\documentclass[a4paper11pt]{article}     
\usepackage[left]{lineno}
\usepackage{color}
\usepackage{amsmath, amscd, amssymb, amsthm, latexsym}

\newcommand{\N}{\mathbb N}
\newcommand{\words}{\{0,1\}^*}
\theoremstyle{plain}

\title{
{\bf\huge  Kolmogorov Complexity in perspective
}
\bigskip
\\{\bf Part II: Classification, Information
Processing and Duality}\footnote{%
Published in Synthese, 2008-2010. A French version is also available 
on the Web.
                                 }%
}
\medskip\medskip
\author{Marie Ferbus-Zanda\\
{\footnotesize LIAFA, CNRS \& Universit\'e Paris Diderot - Paris 7}\\
{\footnotesize Case 7014}\\
{\footnotesize 75205 Paris Cedex 13 France}\\
{\footnotesize Marie.Ferbus@liafa.jussieu.fr}
}
\date{}
\begin{document}
\maketitle
%
%
%
\begin{abstract}
%
\noindent 
We survey diverse approaches to the notion of information: 
from Shannon entropy to Kolmogorov complexity.
Two of the main applications of Kolmogorov complexity are presented:
randomness and classification.
\\
The survey is divided in two parts published in a same volume.
\\
Part II is dedicated to the relation between logic and information system,
within the scope of Kolmogorov algorithmic information theory.
We present a recent application of Kolmogorov complexity:
classification using compression, an idea with provocative
implementation by authors such as Bennett, Vit\'anyi and 
Cilibrasi.  This stresses how Kolmogorov complexity,
besides being a foundation to randomness, 
is also related to classification.
Another approach to classification is also considered:
the so-called ``Google classification''.
It uses another original and attractive idea which is connected
to the classification using compression and to Kolmogorov complexity
from a conceptual point of view. 
We present and unify these different approaches to classification
in terms of Bottom-Up versus Top-Down operational modes,
of which we point the fundamental principles and the underlying 
duality.
We look at the way these two dual modes are used in different approaches
to information system, particularly the relational model for database
introduced by Codd in the 70's.
This allows to point out diverse forms of a fundamental duality.
These operational modes are also reinterpreted in the context of the
comprehension schema of axiomatic set theory ZF.
This leads us to develop how Kolmogorov's complexity
 is linked to intensionality, abstraction, classification 
 and information system.
%
\bigskip
\newpage
\noindent{\bf Keywords:} {Logic, Computer Science, 
Kolmogorov Complexity, Algorithmic Information Theory, 
Compression, Classification, Information System, Database,  
Bottom-Up versus Top-Down Approach,
Intensionality, Abstraction}.
\end{abstract}
%
%
{\footnotesize\tableofcontents}
\normalsize
%
\newpage
%
%
\noindent {\bf Note.}
All notations and definitions relative to
Kolmogorov complexity are introduced 
in Part~I~\footnote{%
Ferbus-Zanda~M. \& Grigorieff~S. 
Kolmogorov Complexity in perspective.
Part~I: Information Theory and Randomness. 
To appear in {\em Synthese}, 2010.
\\One can also consult
\cite{ferbusgrigoBullEATCS2001}, 
\cite{DurandZvonkin-VAA-2004-2007},
\cite{delahaye1999} et \cite{livitanyi}
and the pioneer works of Andrei Nikolaevich Kolmogorov \cite{kolmo65},
Gregory Chaitin  \cite{chaitin66,chaitin75} and
Ray Solomonoff \cite{solo64a,solo64b}.
                   }.%
%
\section{Algorithmic information theory and classification}
\label{s:ThAlgoInfoClassification}
%
\noindent
Using Andrei Nikolaevich Kolmogorov complexity, 
striking results have been obtained
on the problem of classification for quite diverse families of objects:
let them be literary texts, music pieces,
examination scripts (lax supervised) or, at a different level,
natural languages and natural species (phylogeny).
\medskip
\\
The authors, mainly Charles Bennett, Paul Vit\'anyi, 
Rudi Cilibrasi\footnote{%
Jean-Paul Delahaye's surveys \cite{delahaye2004,delahaye2006}
give a clear introduction to these works
(let us acknowledge that they were very helpful for us).
                        }%
~have worked out refined methods which are along the following lines.
%
\subsection{Definition and representation
of the family of objects we want to classify}  
\label{ss:DefFamilleObjets} 
%
First we have to define a specific family of objects 
which we want to classify.
\medskip
\\
For example, a set of Russian literary texts that we want to
group by authors. In this simple case, all texts are written
in their original Russian language.
\medskip
\\
Another instance, music. In that case, 
a common translation is required, i.e., 
a {\em normalization\thinspace}
of music pieces (representing or, in other words,
interpreting musical partitions)
which we want to group by composer.
This common representation (which has to be tailored for
computer programs) is necessary in order to be able to 
compare these diverse music pieces.
Let us cite Delahaye \cite {delahaye2004}:
%
%
\begin{quote}
{\scriptsize$\ll$}~Researchers considered 
36 music pieces coded as MIDI
({\em Musical Instrumental Digital Interface}) files.
They normalized them by producing piano versions
and considering them as data files consisting of long lists
of bytes\footnote{%
A {\em byte\thinspace} 
is a sequence of 8 binary digits. It can also
be seen as a number between 0 and 255.
                  }.%
~Without such a normalization, 
which is a real informations extraction,
nothing would work [\ldots]~{\scriptsize$\gg$}
\end{quote}
%
An instance at a different level: the 52 main European languages.
In that case one has to choose 
a canonical object (here a text) and its representations 
(here translations) in each one of the different languages
(i.e. corpus) that we consider.
For instance, the \textit{\thinspace Universal 
Declaration of Human Rights \thinspace}
and its translations in these languages,
an example which was a basic test for Vit\'anyi's method.
As concerns natural species (another example developed
by Vit\'anyi), the canonical object will be a
DNA sequence.
\medskip
\\ 
{\sl What has to be done is to select, define
and normalize a family
of objects or a corpus that we want to classify.}
\medskip
\\
Normalization of a family of objects is a complex problem,
and it may be also the case for the definition of such a family.
{\em Roughly speaking}, one can partition 
the types of considered objects
in different classes :
%
\begin{itemize}
\item [\textbullet]
Well defined families of objects to be classified.
Normalization of these objects (rather of their representations) can then be 
done without loss of information.
This is the case of literary texts.
\item [\textbullet] 
The family to be classified can be finite though unknown,
possibly without a priori bound on its size.
Such is the case with informations on the Web
(cf.~classification using Google, section~\ref{s:ngd}).  
\item [\textbullet] 
There are some cases where such a normalization is difficult
to work out if not impossible.
It may be the case for painting, drawing, photography, 
art-house cinema, etc.
\end{itemize}
%
\subsection{Comparing the common information content}
\label{ss:ComparaisonContenuInfo} 
%
%
Finally, one gets a family of words in the same alphabet which represent
the objects that we want to compare
and measure the common information content\footnote{%
The notion of information content of an object is detailed in Part~I.
According to Kolmogorov, this is, by definition, 
the algorithmic complexity of that object.
                    }%
~(observe that we can reduce to a binary alphabet).
Our goal  is to compare and, if possible, to measure
the common information content.
\medskip
\\ 
This comparison is done by defining a distance for the pairs of such
(binary) words with the following intuition:
%
\begin{quote}
{\sl
The more common information is shared by two words,
the closer they are and the shorter is their distance.
Conversely, the less common information existing between two words,
the more they are independent and non correlated,
and greater is their distance.
\\
Two identical words have a null distance.
Two totally independent words
(for example, words representing  two events as
$100$ random coin tosses)
have a distance of about $1$
(for a normalized distance bounded by $1$).
}
\end{quote}
%
Observe that the authors, in their approach of classification of 
information, follow  the ideas pioneered 
by Claude Shannon and  Kolmogorov to define
a {\em quantitative\thinspace} measure of information content of 
words, i.e. a measure of their randomness
(in exactly the same way as a volume or a surface gets a numerical
measure).
%
\subsection{Classification}  
\label{ss:Classification} 
%
%
We now have to associate
a classification to the objects or corpus defined in
section~\ref{ss:DefFamilleObjets} using the numerical measures 
based on the distances introduced in 
section~\ref{ss:ComparaisonContenuInfo}.
This step is presently the least formally defined.
The authors give representations of the obtained classifications
using tables, trees, graphs,~etc.
\medskip
\\
This is indeed more a {\em visualization}, 
i.e. a {\em graphic representation},
of the obtained classification than a {\em formal classification}.
Here the authors have no powerful mathematical framework such as
the {\em relational model for databases}~ elaborated by
Edgar~F.~Codd in the 70's \cite{Codd1970,Codd1990}
and its (recent) extension to {\em object database}~ with trees.
Codd's approach is currently one of the sole mathematical formal 
approaches (if not the only one)
to the notion of {\em information structuralization}. 
In this way, one can say that 
structuring a class of informations or 
(representations of) objects (from the ``real world'' as 
computer scientists call it)
amounts to a relational database which is 
itself a perfectly defined mathematical object. 
Moreover one can question this database 
and extract ``new'' informations
via {\em queries} which can be written in a formal 
language (namely {\em Codd's relational algebra\thinspace}).
Also, notice that this extremely original theoretical approach is 
the one which is implemented in all database softwares since the 80's
and is used everywhere there is some mention of databases.
\medskip
\\
{\sl Consequently, the question is how are 
we to interpret in a formal way
tables or trees in classification via compression and  
more particularly how are we
to formally extract informations from this classification?}
\medskip
\\
Though valuable, the classification obtained by this method
(of classification via compression) is rudimentary and
{\em non formal}.
This is somewhat analogous, for instance,
to the classification of words
in a dictionary of synonyms. 
We face a complex problem on which we shall return in
section~\ref{s:ClassifInfoAppBUTD}.      
Nevertheless, Vit\'anyi \& al. obtained by these methods
a classification tree for the 52 European languages which is
the one revealed by linguists, a remarkable success.
They also obtained phylogenetic trees classifying natural
species which are in accordance
with those obtained by paleontologists.
These trees represent parenthood relations between natural species
and are obtained via DNA sequence comparisons.
%
\subsection{Normalization}  
\label{ss:Normalisation} 
%
\noindent
An important problem remains when using a distance 
as in section~\ref{ss:Classification}. 
To obtain a classification, we have to consider the amount 
of information contained in the considered objects.
Let us cite Cilibrasi \cite{cilibrasi03}:
%
\begin{quote}
{\scriptsize$\ll$}~Large objects (in the sense 
of long strings) that differ
by a tiny part are intuitively closer than tiny objects that
differ by the same amount.
For example, two whole mitochondrial genomes of $18,000$ bases
that differ by $9,000$ are very different, while two whole nuclear
genomes of $3\times 10^9$ bases that differ by only $9,000$ bases
are very similar.~{\scriptsize$\gg$}
\end{quote}
%
As we shall see later, this problem is relatively
easy to solve using a normalization of distances. 
Notice that this is a different way of normalization 
that the one proposed in~section~\ref{ss:DefFamilleObjets}.
%
\subsection{Compression}  
\label{ss:Compression} 
%
\noindent
Finally, all these methods rely on Kolmogorov complexity 
which is, as we know, 
a non computable function 
(cf.~for example~\cite{ferbusgrigoBullEATCS2001}). 
\medskip
\\
The remarkable idea introduced by Vit\'anyi
is the following:
\medskip
{\sl
%
\begin{itemize}
\item[\textbullet]
Consider the Kolmogorov complexity of an object
as the ultimate, ideal and optimal value of the compression of 
the representation of that object.
\item[\textbullet]
Compute approximations of this ideal compression
using usual efficient compression algorithms, 
implemented with compressors 
such as gzip, bzip2, PPM, etc.
~which are of common use with computers.
\end{itemize}
%
\noindent
}
Observe that the quality and fastness of such compressors
is largely due to heavy use of statistical tools.
For example, {\em PPM} ({\em Prediction 
by Partial Matching}) uses a pleasing mix of 
statistical\footnote{%
We come back in section~\ref{s:ClassifInfoAppBUTD} below
on information processing with statistics.
                     }%
~models arranged by trees, suffix trees
or suffix arrays.
\medskip
\\
We remark that the efficiency 
of these tools is of course due to
several dozens of years of research in data compression.
And as time goes on, they improve and better approximate
Kolmogorov complexity. 
Replacing the ``pure'' but non computable Kolmogorov complexity
by a banal compression algorithm such as gzip
is quite a daring step taken by Vit\'anyi.
%
\section{Classification via compression}
\label{s:ClassificationCompression}
%
\subsection{The normalized information distance $\text{\em (NID)}$}
\label{ss:nid}
%
We now formalize the notions described above.
The basic idea is to measure the information content shared
by two binary words representing some objects in a family
we want to classify.
\medskip
\\
The first such tentative goes back to the 90's \cite{bennett}:
Bennett and al. define a notion of {\sl information distance}
between two words $x,y$~ as the size of the shortest program
which maps $x$ to $y$~ and ~$y$ to $x$.
This notion relies on the idea of {\sl reversible computation}.
A possible formal definition for such a distance is
$$
\text{\em ID'}\thinspace(x,y) = \mbox{least 
$|p|$ such that $U(p,x)=y$ and $U(p,y)=x$}
$$
where ~$U:\words\times\words\to\words$ is optimal for 
the conditional complexity $K(\ |\ )$ (cf.~Part~I).
\\
We shall mainly work with the following alternative definition: 
$$
\text{\em ID}\thinspace(x,y) = \max\{K(x|y),K(y|x)\}
$$
{\sl The intuition for these definitions is that the shortest program
which computes $x$ from $y$ and $y$ from $x$
takes into account all similarities between $x$ and $y$}.
\medskip
\\
Observe that the two definitions do not coincide
(even up to logarithmic terms) but lead to similar
developments and efficient applications.
%
%
\medskip
\\
{\bf Note.}
In the definition of $\text{\em ID}$, we can consider $K$ 
to be plain Kolmogorov complexity
or its prefix version (denoted $H$ below).
In fact, this does not matter for a simple reason:
all properties involving this distance will be true up to a
$O(\log(|x|),\log(|y|))$ term and the difference between
$K(z|t)$ and $H(z|t)$ is bounded by $2\log(|z|)$.
For conceptual simplicity, we stick to plain Kolmogorov complexity.
%
%
\medskip
\\
$\text{\em ID'~}$ and $~\text{\em ID}~$
satisfy the axioms of a distance
{\em up to a logarithmic term}.
\medskip
\\The strict axioms for a distance $d$ are
$$
\left\{\begin{array}{rcll}
d(x,x)&=&0&\mbox{~~~~{\em(identity)}}\\
d(x,y)&=&d(y,x)&\mbox{~~~~{\em(symmetry)}}\\
d(x,z)&\leq&d(x,y)+d(y,z)&\mbox{~~~~{\em(triangle inequality)}}
\end{array}\right.
$$
\medskip
\mbox{}\\
{\bf Theorem.}
\\
{\em The up to a $\log$ term distance
axioms which are satisfied by ~$\text{\em ID'}~$
and ~~$\text{\em ID}$~ are as follows:
$$
\left\{\begin{array}{rclr}
d(x,x)&=&O(1) &~~~~(1)\\
d(x,y)&=&d(y,x) &~~~~(2)\\
d(x,z)&\leq
&d(x,y)+d(y,z) + O(\log(d(x,y)+d(y,z))) &~~~~(3)
\end{array}\right.
$$}
%
%
\medskip
{\small\begin{quote}
\begin{proof}
We only treat the case of $\text{\em ID}$.
Let $f:\words\times\words\to\words$ be such that
$f(p,x)=x$ for all $p,x$.
The invariance theorem insures that $K(x|x) \leq K_f(x|x)+O(1)$.
Now, taking $p$ to be the empty word, we see that $K_f(x|x)=0$.
Thus, $\text{\em ID}(x,x) = O(1)$.
\\
Equality $\text{\em ID}(x,y)=\text{\em ID}(y,x)$ is obvious.
\\
Let now $p,p',q,q'$ be shortest programs such that
$U(p,y)=x$, $U(p',x)=y$, $U(q,z)=y$, $U(q',y)=z$.
Thus,
$K(x|y)=|p|$, $K(y|x)=|p'|$, $K(y|z)=|q|$, $K(z|y)=|q'|$.
Consider the injective computable function
$\langle\ \rangle:\words\times\words\to\words$
(cf.~Proposition 1.6 in Part~I)  which is such that
$|\langle r,s \rangle|=|r|+|s|+O(\log |r|)$.
Let $\varphi:\words\times\words\to\words$ be such that
$\varphi(\langle r,s \rangle, x) = U(s,U(r,x))$.
Then
$$
\begin{array}{rclclcl}
\varphi(\langle q,p \rangle,z) &=& U(p,U(q,z)) &=& U(p,y) &=& x\\
\varphi(\langle p',q' \rangle,x) &=& U(q',U(p',x)) &=& U(q',y) &=& z
\end{array}
$$
so that, applying the invariance theorem, we get
%
\begin{center}
\begin{tabular}{ll}
$K(x|z)$ & $\leq K_\varphi(x|z)+O(1)\leq|\langle q,p \rangle|+O(1)$
\\
       & $= |q|+|p|+ O(\log(|q|))= K(y|z)+K(x|y)+O(\log(K(y|z)))$
\end{tabular}
\end{center}
%
and, similarly, $K(z|x) \leq K(y|x)+K(z|y)+O(\log(K(z|y)))$.
Thus,
%
\begin{eqnarray*}
\max(K(x|z),K(z|x)) &\leq& \max(K(y|z)+K(x|y)+O(\log(K(y|z))),\\
&&\phantom{\max(} K(y|x)+K(z|y)+O(\log(K(z|y))))
\\
&\leq&
\max(K(x|y),K(y|x))+ \max(K(y|z),K(z|y))\\
&& \qquad \qquad +O(\log(\max(K(y|z),K(z|y))))
\end{eqnarray*}
%
Which means
$\text{\em ID}(x,z) \leq \text{\em ID}(x,y)
                                        + \text{\em ID}(y,z)
                                        +O(\log(\text{\em ID}(y,z)))$,
a slightly stronger result than (3).
\end{proof}
\end{quote}}
%
\medskip
\noindent
It turns out that such approximations of the axioms
are enough for the development of the theory.
\\
To avoid scale distortion, as said in 
section~\ref{ss:Normalisation},
distance $\text{\em ID}$ is normalized to
$\text{\em NID}$
({\em normalized information distance\thinspace}) as follows:
$$
\text{\em NID\thinspace}(x,y) = 
\frac{\text{\em ID\thinspace}(x,y)}{\max(K(x),K(y))}
$$
The remaining problem is that this distance is not computable
since $K$ is not.
Here comes Vit\'anyi's daring idea:
%
\begin{quote}
{\sl
Consider \thinspace$\text{\em {NID}}$\enskip as an ideal distance which
is to be approximated by replacing the Kolmogorov function $K$
by computable approximations obtained via compression algorithms.
}
\end{quote} 
%
\subsection{The normalized compression 
distance $\text{\em (NCD)}$}
\label{ss:NCD}  
%
%
The approximation of $K(x)$ by $C(x)$ where $\Gamma$ is a 
compressor\footnote{%
A formal definition of compressors is given in Part~I.
                     },%
~does not suffice. We also have to approximate the conditional
Kolmogorov complexity $K(x|y)$.
Vitanyi chooses the following approximation:
$$
\Gamma(y|x) = \Gamma(xy) - \Gamma(x)
$$
The authors explain as follows their intuition:
\medskip
\\
{\sl To compress the word $xy$ ($x$ concatenated to $y$)
%
\begin{itemize}
\item [\textbullet] 
The compressor first compresses $x$.
\item [\textbullet] 
Then it compresses $y$ but skips all information from $y$
which was already in $x$.
\end{itemize}
%
Thus, the output is not a compression of $y$ but a compression
of $y$ with all $x$ information removed,
i.e. this output is a {\em conditional compression\thinspace} 
of $y$ knowing $x$.
}
\medskip
\\ 
Now, the assumption that in the compression of the word $xy$
the compressor first compresses $x$ is questionable: 
how does the compressor recovers $x$ in $xy$?
One can argue positively in case $x$ and $y$ are random
(i.e. incompressible) and in trivial case $x=y$.
And between these two extreme cases? But it works.
The miracle of modeling?
Or something not completely understood?
\medskip
\\
With this approximation, plus the assumption that
$\Gamma(xy)=\Gamma(yx)$ (also questionable: it depends on the used compressor)
we get the following approximation
of $\text{\em {NID}}$, called the 
{\em normalized compression distance}, 
$\text{\em NCD}$ :
\begin{eqnarray*}
{\mbox{\em NCD\thinspace}}(x,y) &=& 
\frac{\max~(\Gamma(x|y)~,~\Gamma(y|x))}{\max~(\Gamma(x)~,~\Gamma(y))}\\
&=& \frac{\max~(\Gamma(yx)-\Gamma(y)~,~\Gamma(xy)-\Gamma(x))}
{\max~(\Gamma(x)~,~\Gamma(y))}\\
&=& \frac{\Gamma(xy)-\min~(\Gamma(x)~,~\Gamma(y))}
{\max~(\Gamma(x)~,~\Gamma(y))}
\end{eqnarray*}
\medskip
\\
Remark that clustering according to $\text{\em NCD}$ and, 
more generally, classification
via compression, is a 
{\em black box\thinspace}\footnote{%
\label{BoiteNoire} 
The notion of black box is a scientific 
concept introduced by Norbert Wiener in 1948:
%
Wiener~N.
{\em Cybernetics or Control and Communication 
in the Animal and the Machine}.
The Technology Press, 1948 \& 2nd Ed. The MIT Press, 1965.
%
This concept is one of the fundamental principles of Cybernetics.
It is issued from the multidisciplinary 
exchanges during the Macy conferences
which were held in New-York, 1942--1953.
Indeed, the emergence of cybernetics and information theory
owes much to these conferences.
                                  }%
~as noticed by Delahaye
\cite{delahaye2006}: words are grouped
together according to features that are not explicitly known to us
except if we had already a previous idea.
Moreover, there is no reasonable hope that the analysis of the
computation done by the compressor gives some light on the obtained
clusters.
\medskip
\\
For example, what makes a text by Tolstoi so characteristic?
What differentiates the {\em styles\thinspace}  
of Tolstoi and Dostoievski?
But it works, Russian texts are grouped by authors by a compressor
which ignores everything about Russian literature\ldots
\medskip
\\
When dealing with some classification obtained by compression,
one should have some idea about this classification:
this is semantics whereas the compressor is purely syntactical
and does not ``understand'' anything''. Thus one cannot hope
some help in understanding (interpretation) of
the obtained classification
(cf.~section~\ref{s:ClassifInfoAppBUTD}).
This is very much as with machines which, given some formal
deduction system, are able to prove quite complex statements.
But these theorems are proved with no explicit semantical idea,
how are we to interpret them? No hope that the machine gives
any hint, at least in the present context.
%
\section{The Google classification}
\label{s:ngd}
%
%
\noindent
Though {\em stricto sensu}, it does not use 
Kolmogorov complexity, we now present
another recent approach by Vit\'anyi \& Cilibrasi \cite{cilibrasi07}
to classification which leads to a very performing tool.
%
\subsection{The normalized Google distance $\text{\em {(NGD)}}$}
\label{ss:ngd}
%
\noindent
This quite original method is based on the huge data mass,
constituted by the Web and which is accessible 
with search engines as Google. 
They allow for basic queries using a simple keyword or
conjunction of keywords.
Observe that the Web (the {\em World Wide Web\thinspace})
is not a formal database:  
it is merely a crude data bank, in fact
a (gigantic) informal information system
since data on the Web are not structured 
as data in relational database.
It has a rudimentary form of structuralization based on graphs and
{\em graphical user interfaces}. Nevertheless, it is endowed with an
object-oriented programming language, namely, Java.
What is remarkable is that there exists 
a norm for this programming language
and, moreover, this language is Turing-complete 
(cf.~section~\ref{ss:SIBDApprocheFormelle}).
This can explain the success
(and fashion) of Java and of the  
{\em object approach\thinspace} which is also 
largely due to the success of the Web.
\medskip
\\
Citing Alberto Evangelista et Bjorn Kjos-Hanssen 
\cite{evangelista06}, the idea of the 
method is as follows:
%
\begin{quote}
{\scriptsize$\ll$}
When the Google search engine is used to search for the word $x$,
Google displays the number of hits that word $x$ has.
The ratio of this number to the total number of Web pages indexed
by Google represents the probability that word $x$ 
appears on a Web page [...]
If word $y$ has a higher conditional probability to appear on a Web
page, given that word $x$ also appears on that Web page, than it does
by itself, then it can be concluded that words $x$ and $y$ are
related.~{\scriptsize$\gg$}
%
\end{quote}
Let us cite an example from Cilibrasi and Vitany \cite{cilibrasi05},
which we complete with updated 
figures\footnote{%
Point~4, section \ref{ss:DiscussionMethode}
relativizes the obtained results.
             }.%
~The searches for the index term ``horse", ``rider" and ``molecule"
respectively return $156$, $62.2$ and $45.6$ million hits.
Searches for pairs of words ``horse rider" and ``horse molecule"
respectively return $2.66$ and $1.52$ million hits.
These figures stress a stronger relation between the words
``horse" and ``rider" than between ``horse" and ``molecule".
\medskip
\\ 
Another example with famous paintings:
``Le d\'ejeuner sur l'Herbe",``Le Moulin de la 
Galette" and ``La Joconde".
Let refer them by a, b, c.
Google searches for a, b, c respectively give
$446~000$, $278~000$ and $1~310~000$ hits.
As for the searches for the conjunctions  
a+b, a+c and b+c, they respectively give
$13~700$, $888$ and $603$ hits. 
Clearly, Jean Renoir's paintings are more often cited 
together than each one is with Leonardo da Vinci's paintings.
\\
In this way, the method regroups paintings by artists, using what
is said about these paintings on the Web.
But this does not associate the painters to groups of paintings
(we have to add them ``by hand'').
\medskip
\\
Formally, Cilibrasi and Vitany \cite{cilibrasi05,cilibrasi07}
define the {\em normalized Google distance\thinspace} as follows:
$$
\text{\em {NGD\thinspace}}(x,y) =\frac{\max
(\log \Lambda (x),\log \Lambda(y))-\log \Lambda(x,y)}
{\log \Upsilon -\min(\log \Lambda(x),\log \Lambda(y))}
$$
where ~$\Lambda(z_1,...z_n)$~  is the number of hits for
the conjunctive query 
$z_1\wedge ... \wedge z_n$ (which is $z_1 ~...~  z_n$ with Google;
If $ n = 1$, $\Lambda(z)$ is the total number of hits for
the query $z$).
$\Upsilon$ is the total number of Web pages that Google indexes.
%
\subsection{Discussing the method}
\label{ss:DiscussionMethode}
%
%
Let us cite some points relative to the use of such a classification 
method (the list is not exhaustive):
%
%
\medskip
\\
1) The number of objects in a future classification and that of
canonical representatives of the different corpora is not chosen
in advance nor even boundable in advance and it is constantly
moving.
This dynamical and uncontrolled feature
of a definition of a family is a totally new
experience, at least for a formal approach of classification.
%
%
\medskip
\\
2) Domains a priori completely rebel to classification such as
the pictorial domain a priori no normalization of 
paintings being possible or if it is this is not obvious 
in the present context can now be easily considered.
This is also the case (and for the same reasons) for sculpture,
architecture, photography, art-house cinema, etc.
This is so because we are no more dealing with the works themselves
but with a discourse about them (which is the one on the Web).
This speech depends on a ``real'' language: a natural 
language or a formal language. Notice that the notion of 
``pictorial language'' remains a metaphor 
as long as we consider that infra verbal communication 
is not a language in the usual sense.
\\
{\sl The discourse which is considered by Google
is the one of the keywords and relations between them, these
keywords coming from queries proposed for the $\text{{NGD}}$ and
appearing in the texts of the users of the Web.}
\medskip
\\
Notice that there are some works 
which can be used for an algorithmic approach 
(possibly a normalization) of pictural pieces, art-house films, etc.
For instance, the French psychoanalyst Murielle Gagnebin elaborated
a theory of {\em esthetics\thinspace} and {\em creation}, based on 
psychoanalysis and philosophy. 
Her meta psychological model is quite efficient to point out the
fundamental psychical mechanisms involved in art pieces.
And this is done from the art pieces themselves, excluding any
discursive consideration on these art pieces or on the artists
Such a model could much probably be implemented as an expert system.
%
%
\medskip
\\
3) However, there is a big limitation to the method, 
namely that one which is called:
the {\em closed world assumption}. 
That can be interpreted as follow:
{\sl the world according to 
Google\thinspace\footnote{%
Irving J. {\em The World According to Garp}. Modern Library, 1978.
                          },%
~information according to Google}, etc.
\medskip
\\ 
If Google finds something, how can 
one check its {\em pertinence}.
Else, what does it mean? 
How can we define (in a general manner) a notion of pertinence
for the informations found by Google?
Sole certainty, that of uncertainty!
Moreover, we notice that
when failing to get hits with several keywords, we give up the
original query and modify (we change its semantics)
it up to the point Google gives some
``pertinent'' answers.
That sort of failure is similar to the use of negation in the
Prolog programming language (called {\em negation as failure}), 
which is much weaker
than classical negation and which is connected to  
the closed world assumption for databases.
\medskip
\\
When failing to get hits, it is reasonable to give up the query and
accordingly consider the related conjunction as meaningless. 
However, one should keep
in mind that this is relative to the closed,
and relatively small, world of data on the Web,
the sole world accessible to Google. Also one 
has not to underestimate the changing aspect of the informations
available on the Web.
When succeeding with a query, the risk is to stop on this
succeeding query and
%
\begin{itemize}
\item [\textbullet] 
Forget that previous queries have been tried and have failed.
\item [\textbullet]
Omit going on with some other queries which could possibly
lead to more pertinent answers.
\item [\textbullet]
Given a query, the answers obtained from Google are those
found at a given moment in a kind of
{\em snapshot} of the Web.
But such an instantaneous snapshot betrays
what is the essence of  the Web: to be a
continuously moving information system.
All the updates (insertions, deletions,
corrections, etc.) are done in a massively parallel
context since Google uses about 700~000 computers
as servers!
Thus, Google answers to a query are not at all
{\em final} answers
nor do they constitute {\em an absolute answer}.
And this is in contrast with the perfect determinism
we are used when computer programs are run
(in this way, Prolog is considerably more 
``deterministic'' than Google) or with databases (when they are
well written...)
Also, the diverse answers given by Google may contradict 
one another, depending on the sites Google retained.
In particular, one is tempted to stop when
a site is found that gives an answer which seems 
convenient (indeed, this is what we do in most cases).
\end{itemize}
%
%
4) 
So we see some difficulties emerging with the theoretical approach
to how Google deals with information from the Web
(and the same for any browser).
For such a reflection, we have chosen an idealistic perspective
where Google searches according to scientific criteria or at least
with some transparency (in particular, on how Web pages are indexed,
or even how many are really indexed).
However let us mention that there are some controversies about 
the indexing and consequently 
on exactness of the results found by Google,
in particular, about the number of
occurrences of a given word of all existing Web pages
(even if not dealing with the content of these pages).
Indeed, some queries lead to very surprising results:
``Googlean logic" is quite strange (when compared with Boolean logic).
This is shown in a very striking (and completely scientific) manner
by Jean V\'eronis in his
blog\footnote{%
V\'eronis~J.
Web~: Google perd la boole. (Transl: Web: Googlean logic.)
Blog. January 19, 2005 from
http://aixtal.blogspot.com/2005/01/web-google-perd-la-boole.html~.
\\See also: http://aixtal.blogspot.com/2005/02/web-le-mystre-des-pages-manquantes-de.html
\\and~  http://aixtal.blogspot.com/2005/03/
google-5-milliards-de-sont-partis-en.html~, 2005.
                }.%
\medskip
\\
A highly important task still to be done is to formalize
the notion of information on the Web
and the relations ruling the data it contains, as
it has been done by Codd with the relational model 
for databases in the 70's.
Previous to Codd's work, organizing and structuring 
data and information in a computer and their accessibility
via the notion of query was underlaid by no solid mathematical 
foundation and was resting on technical tricks. This is still the 
case for the data on the Web.
This remarkable innovative approach via Google search
is still in its infancy.
%
%
\medskip
\\
In the next sections, we consider some formalized notions
together with not yet formalized ideas
--~such as those pointed out in~\ref{ss:Classification}~--
Ongoing work in progress
and some papers are in preparation\footnote{%
In particular
%
\cite{FerbusZanda--Article--DualiteLogComputerBooleanAlg},
\cite{FerbusZanda--Article--LogSICerveauPsy} et
Ferbus-Zanda~M.
{\em Logic, Information System and Metamorphosis of a 
Fundamental Duality}.
In preparation.    
%
                            }.%
%
\section{Classification, Bottom-Up versus Top-Down approaches
and duality}
\label{s:ClassifInfoAppBUTD}    
%
\subsection{Bottom-Up versus Top-Down modes}  
\label{ss:BottomUpvsTopDown}   
%
\noindent
These approaches to classification via compression
and Google search (relative to information appearing 
on the Web in the second case),
are incredibly original and present a huge interest.
With the phenomenal expansion of computer science, nets and the Web,
information has a kind of new status.
So that these approaches
(which are indeed based on what they are able to make explicit)
help us to grasp this entirely new status of information
as it is now with such a world of machines.
\medskip
\\
Whereas classification via the relational model for databases
has a neat formalization, we have stressed above how difficult it is
to formally define the classification obtained by compression or
via Google.
Of course, one could base a such formalization on trees and graphs.
But with such structures, the way information is recovered is rather poorly
formalized.
This is in fact what happens with the organization of files in 
an {\em operating system\thinspace} since none of them uses any database
(let it be {\em Unix, Linux, MacOs, Windows\thinspace} 
and their variants).
\medskip
\\
It seems to us that one should reconsider these different approaches to the 
notion of classification in terms of {\sl two fundamental modes
to define mathematical and computer science objects which are also 
found in the execution of computer programs}.
These two main approaches to define 
mathematical and computer science objects are:
%
\begin{itemize}
\item [\textbullet] {\em Iterative definitions} 
(based on {\em set theoretical union\thinspace})
\item [\textbullet] {\em Inductive (or recursive) definitions}
(based  on {\em set theoretical intersection\thinspace}).
\end{itemize}
%
For instance, one can define propositional formulas,
terms and first-order logic formulas following any one of these two ways.
\medskip\\
Recall that Stephen Kleene's 
presentation\footnote{%
Kleene formally and completely characterizes the notion of
{\em recursive function\thinspace}
(also called {\em computable function\thinspace}), 
by adding the minimization schema 
(1936) to the composition and recursion schemas --~these two last
schemas characterize the {\em primitive recursive functions\thinspace}
which constitute a proper subclass of the class of computable functions:
the {\em Ackermann function\thinspace} (1928) is computable but not
primitive recursive.
From a programming point of view, the minimization schema corresponds
to the {\tt while} loop ({\tt while F(x) do P} where {\tt F(x)} is a Boolean valued
property and {\tt P} is a program).
Cf. also Note~\ref{TheseChurchTuring} or the book by Shoenfield~J.
{\em Recursion theory}, Lectures Notes in Logic, 
(new edition) 2001.
                   }%
~of partial recursive 
functions is based on three
(meta) operations: {\em composition\thinspace}, 
{\em primitive recursion\thinspace} and {\em 
minimization}. 
%
\begin{itemize}
\item [\textbullet] 
\medskip
Iterative definitions are 
connected to minimization 
(and to the notion of {\em successor\thinspace}). We can describe
these type of definitions as ``{\em Bottom-Up\thinspace}'' 
characterizations.
\item [\textbullet] 
Inductive definitions are connected to primitive 
recursion (and to the notion of 
{\em predecessor\thinspace}). 
We can describe
these type of definitions as ``{\em Top-Down\thinspace}'' 
characterizations.
\end{itemize}
%
Notice that
composition is related to both characterizations, the
bottom-up and top-down ones.
We gave, in Part~I, formalizations of randomness for infinite
objects which follow these two bottom-up and top-down approaches
(cf.~Part~I, section~5.1 and 5.2).
These two modes are also found in the execution of computer programs:
%
\begin{itemize}
\item [\textbullet] 
{\em Execution in the iterative mode\thinspace} is called 
{\em Bottom-Up\thinspace}.
\item [\textbullet]
{\em Execution in the recursive  mode\thinspace} is called
{\em Top-Down\thinspace}.
\end{itemize}
%
This last mode requires the use of a {\em stack}
which goes on growing and decreasing
and into which results of intermediate computations are pushed
until getting to the ``basic cases'', 
i.e. the initial steps of the inductive definition
of the program which is executed.
To execute an iterative program, 
all data necessary for its execution
are at disposal without need of any stack.
From the computer scientist point of view,
these two execution modes
are really far apart.
Notice that the execution mode (iterative or recursive) follows the
definition mode (iterative or recursive) of the program to be executed.
Nevertheless, in some cases, recursive programs may be executed in an 
iterative way avoiding any 
stack\footnote{%
This is the case for {\em tail-recursion\thinspace}
definitions. In some programming 
languages such as {\em LISP\thinspace} 
such tail-recursion programs are 
generally executed (when the programs executor is well 
written) in an iterative way. 
Tail-recursion programs represent a {\em limit case\thinspace} 
between iterative programs and recursive programs.
               }.%
\medskip
\\
In the same way, one observes that 
there are ~{\em two  modes\thinspace}
--~let us also call them {\em Bottom-Up} and {\em Top-Down}~--
that are used in the approach to classification 
of information and/or objects
(of the real world) which are formally represented as words
or more generally as 
texts\footnote{%
Depending on how much {\em abstraction\thinspace} is wanted 
(or how much {\em refinement\thinspace}  is
wanted), a {\em text\thinspace} will be represented 
by a {\em binary word\thinspace}
(the blank spaces separating words being also encoded as special characters)
or by a {\em sequence of binary strings\thinspace} 
(each word in the text being represented
by a string in the sequence).
It is also possible to consider sequences or sets of texts and to mix 
such sequences and/or sets
In this paper, we mostly consider encodings of texts with binary words
(in particular, for the examples) and not sequences of binary words,
and we consider sets of such texts.
                }%
~or even as sets of words,
in some alphabet (which can, as usual, be supposed to be binary).
%
\begin{itemize}
\item [\textbullet] 
{\sl In the Bottom-Up mode, one enters into information details.
Otherwise said, one accesses the content of texts, i.e. the 
words representing the diverse informations and/or objects that
one wants to classify (and the meaning of these words and/or texts). 
Texts, families of words, etc. are grasped 
from the {\em inside\thinspace} and their meaning
is essential}.
\item [\textbullet]
{\sl In the Top-Down mode, one does not access the content of texts
in the above way. Texts are, in fact, handled from the {\em outside},
that is ``from the 
top and down"}\thinspace\footnote{%
It is one way of seeing things! The one reflected by
the Anglo-Saxon terminology "top-down".
What is essential is that texts are apprehended 
from the {\em outside}, 
in opposition to apprehension from the {\em inside}.
                         }.%
~{\sl To say things otherwise, one uses a kind of ``oracle'' to grasp
texts and families of words, i.e. means that are {\em exterior\thinspace}
to the the understanding of text and the content of words.}
\end{itemize}
%
Let us illustrate this with an example: the use of keywords to
structure families of texts.
One then uses both bottom-up and top-down modes to classify texts
in the following way: 
%
%
\medskip
\\
1)
It is usual to follow a bottom-up approach in the choice of keywords.
Particular words in texts are
{\sl chosen in consideration of the content of texts and their meaning}
and in order to facilitate future searches.
More precisely, some words will be considered as keywords and will
be declared as.
This is typically the case with scientific papers where keywords 
are chosen by the author, the journal editor, the librarian, etc. in
view of future classification.
Of course, this supposes that the texts 
have already been read (and understood).
\medskip
\\
Observe that translating a text 
into a natural language to another one
(as, for example, this paper from French to English)
requires such a reading and (subtle) understanding of the 
text\footnote{%
With a purely syntactic automatic translator,
such as the one in Google, one can get results
like the following one:
``Alonzo Church" translated as ``\'Eglise d'Alonzo"
(i.e. church in Alonzo)!
               }.%
\medskip
\\
One can also choose keywords for a text 
using totally different criteria.
For instance, rather than reading  the text itself,
one can read and understand an outline or the table of contents
and this is also a bottom-up mode.
One can also look at an index 
(if it exists some): a {\em limit case}
which follows a top-down mode. Indeed, no understanding of the words
is required to select keywords
(though, of course, it does not harm to understand them),
one only consider which words
are mentioned  in the index and their relative importance
(which a good index makes clear).
Without index, one can also count occurrences of words in a text:
this is precisely what Google does in its searches.
In practice, both bottom-up and top-down modes are often used together
({\em mix\thinspace} mode).
\medskip
\\
Whatever method is chosen, the choice of keywords generally
assumes (though it is not always the case) a preliminary knowledge
or some general idea of the wanted classification
for which the keywords are chosen.
This knowledge is a very abstract form of 
{\em semantics\thinspace}\footnote{%
\label{note:Thesaurus}
Let us mention that a new concept emerged: that of
{\em thesaurus\thinspace} which is somehow an abstract semantics
related to classification.
A thesaurus is a particular type of {\em documentary language\thinspace}
(yet a new concept) which, for a given domain, lists along a graph
words and their different relations:
{\em synonymy}, {\em metaphor}, {\em hierarchy}, 
{\em analogy}, {\em comparison}, etc.
Thus, a thesaurus is a kind of {\em  normalized 
and classified vocabulary\thinspace} for a particular domain.
Otherwise said, the words a thesaurus contains,
constitute an {\em hierarchical 
dictionary\thinspace}  of keywords for the considered domain.
One can possibly add definitions of words or consider the classification
of words (according to the future usage of the thesaurus) to be sufficient.
It is a remarkable tool. First used for disciplines around documentation
and for large databanks, it is now used almost everywhere.
To build a thesaurus, one follows a bottom-up or a top-down mode or
mixes both modes, exactly like in the case of keywords.
More details on the notion of thesaurus
in the section devoted to databases
(cf.~section~\ref{ss:SIBDApprocheFormelle} ).
                     },%
~which can evolve through time as new texts are being read.
Generally, the person who writes the text is not the one
who has this knowledge, this is rather the person 
who ``manages" the classification.
\medskip
\\ 
2)
Whatever approach was used, once the keywords have been chosen
and stored in some way, they give a kind of classification for this
text considered among the other ones which have been treated
in a similar way.
Using given keywords, one can look for all texts which have been
assigned such keywords.
Clearly, a notion of {\em query} emerges from the so used keywords.
In an extended concept of keyword
--~and this is exactly how Google works~--
one can look for all texts containing these keywords
(i.e. with these keywords in their contents), with no need
to define any keyword for texts.
\medskip
\\
Observe that searching using keywords 
--~and this is a fundamental point~--
{\sl is a top-down approach to texts
and to their classification}.
A set of keywords (a ``conjunction" of keywords) 
is a form of question to some {\em oracle},
so as to grasp texts from the outside,
without reading nor understanding them.
Using such a set of keywords, one can select some texts among
a family of texts which can be really big, even gigantic in the case
of the Web.
The selected texts can then possibly be read and be understood
(at least better understood).
One can also group them with other texts having some common keywords
and thus get a classification of texts.
\medskip
\\
3)
With the Google approach to classification, things are similar:
the choice of keywords for queries to Google
(which are in fact conjunctions of keywords)
can be done in two ways.
%
\begin{itemize}
\medskip
\item [\textbullet]
In a bottom-up mode this choice comes from the reading and
understanding of the content of the Web.
\item [\textbullet] 
In a top-down mode this choice is based on criteria totally
exterior to the content of the Web though it is hard not to be
somewhat influenced by previous readings from the Web\ldots
\end{itemize}
%
In general, both bottom-up and top-down modes are mixed for
the choice of keywords.
\medskip
\\
Whatever approach was used, once the keywords have been chosen,
one has at disposal a kind of oracle to grasp the Web.
Otherwise said, {\sl the Google query written with these keywords
will select texts from the Web --~and also hypertexts or multimedia data:
pages from the Web~-- with a top-down operational mode}.
Such selected texts can then be read, classified, etc.
Thus, Google as a web search engine behaves as an {\em oracle},
which given some query (keywords), returns a set of web sites.
The way Google works, as for the oracles, is totally 
invisible to the user.
\medskip
\\
One can also surf the Web along a bottom-up mode,
that is give up query and go from one page to another one via
the {\em links hypertext}.
Indeed, those links are the main originality of the Web.
From a theoretical point of view, they are very interesting since they
convey a form of semantics. 
Thus, the {\sl notion of keywords (and more generally of words)
appears to be a limit concept between syntax and semantics}. 
In general {\sl surfing is done via both approaches:  
bottom-up with hypertext links and top-down with queries.}
As before, the choice of the keywords submitted to Google
in view of a classification is also a form of semantics.
Observe that in a top-down approach for the choice of keywords,
one can however choose them {\em randomly\thinspace}, then
use some counting (with statistical tools) to get classifications
of the selected texts.
Such random choices are particularly interesting when there is
a huge quantity of texts, which is the case with the Web.
However, it is doubtful that such an approach to classification 
--~if fundamentally random~-- can give significant results.
Nevertheless, it can be coupled with a more ``deterministic" approach.
\medskip
\\
Let us go back to the title of this section:
{\sl Bottom-Up versus Top-Down modes}. It is reasonable to question:
why are there two possible modes in the definition of mathematical 
and computer science objects and in the runs of computer programs?
{\em De Facto}, these two modes do exist and they are the fundamental
modes which have emerged from the works of the diverse researchers in
computability theory in the XX~th century.
We have seen that these two modes could also be considered in the approach
to classification of information and we gave an example with keywords.
We have also seen how the use of Google to search the Web was relevant to
these approaches.
\medskip
\\
This shows that these bottom-up and top-down modes are not particular to
classification: they concern, in fact, any information processing hence any,
more or less abstract theory of information.
This also concerns all disciplines which deal in some way with the notion 
of {\em representation\thinspace} or {\em definition\thinspace},
or {\em description\thinspace}, etc.
This includes logic, Kolmogorov complexity and computer science,
semiotics and also all sciences of cognition: as we detail in
\cite{FerbusZanda--Article--LogSICerveauPsy},
{\sl  information processing by the human 
brain could fundamentally be structured
around these two operational modes}.
In any case, this is quite an interesting approach to 
cognition which is much enlightened
by the evolution of mathematical logic and computer science.
\medskip
\\
In this paper, we shall look at these 
bottom-up and top-down modes in two
types of situations (concerning classification)
which generalize what we said about keywords.
Namely,
%
\begin{itemize}
\item [\textbullet] 
The logical formalization of information systems via databases
(section~\ref{ss:SIBDApprocheFormelle}).
\item [\textbullet]
The set theoretical approach to the notion of grouping, 
based on the Zermelo-Fraenkel (ZF)
axiomatic set theory.
We shall particularly look at the comprehension schema in ZF
(section~\ref{s:InterpEnsDualiteBUvsTD}).
\end{itemize}
%
%
\medskip
These reflections will help to understand the role played 
by the Kolmogorov complexity in information classification 
and more precisely in the notion of {\em grouping\thinspace} 
of informations. We will have to reconsider the notions of 
intensionality, abstraction semantics and representation in this context
(cf~section~\ref{s:InformationIntentionnaliteK}).
\medskip
\\
Also notice that the existence of 
two such modes for the definitions of mathematical 
and computer science objects, functions and programs and 
for the execution of these programs, is quite interesting. 
The fact that we find these two modes for the various forms 
of the information processing and different disciplines,
of the information processing indicates that this observation 
is a fascinating scientific project. Clearly these two modes, so 
{\em complementary\thinspace}, form a {\em duality 
relation\thinspace}, a kind of correspondence  between
two distinct ways of processing which are somewhat distinct and also
similar\footnote{%
The abstract notion of {\em isomorphism \thinspace} in mathematics 
is a form of duality. Some dualities are not reduced to isomorphisms.
Typically, Boolean algebras with the complement operation (in addition 
to additive and multiplicative operations) contain an internal 
duality and are the basis of deep dualities such as Stone duality 
which links the Boolean algebras family and some topological spaces.
The complement operation confronts us to many problems
and deap results\ldots
                   }.%
~More precisely, we have seen that the bottom-up approach 
(on which are based the iterative definitions), 
results from the notion of set theoretical union
whereas the top-down approach 
(on which are based the inductive definitions),
results from the notion of set theoretical intersection.  
It is therefore quite natural to revisit these approaches 
in the framework of Boolean algebras, a theory 
where the notion of duality is typical, so we do in   
\cite{FerbusZanda--Article--DualiteLogComputerBooleanAlg}.
\medskip
\\ 
Other fundamental dualities for logic and 
computer science are also developed in those papers. 
Especially, duality {\em syntax versus semantics}
and also duality {\em functional versus relational
\thinspace}\footnote{%
Since Gottlob Frege invention, at the end of the 
19th century, of the mathematical logic and the 
formalization of the  mathematical language
that results from it, mathematicians have {\em de facto}
to deal with two distinct categories of mathematical 
symbolsÊ: the {\em function symbols\thinspace} and  
{\em relational symbols\thinspace} 
(or predicate symbols) in complement
of symbols representing objects. To each of these
two large classes of symbols respectively 
correspond algorithms and information systems.
                     }%
~which concerns, among others, the relation between
algorithms and (functional) programming on the one hand
and discrete information system and their 
formalizations\footnote{%
The information systems in which we highlight a
type of programming that we named 
{\em relational programming\thinspace} in a research report:
\label{cite:RapportRech-LITP-ResolProlog} 
Ferbus-Zanda~M.
{\em La m\'ethode de r\'esolution et le langage Prolog\thinspace}
({\em The resolution method and the language Prolog}\thinspace).
Rapport LITP, No-8676, 1986.
We present in this paper the link between functional 
programming and relational programming.
                }%
~on the other hand.
Recall that the essential part of discrete information 
system is the organization (the structuralization) of information, 
whatever is their nature (admitting a discrete representation), 
with the objective of an easily extracting particular informations. 
Clearly, information system are linked to classification. 
Thus, we believe it is interesting to present them 
in this paper. We shall articulate this presentation around 
the bottom-up versuss top-down duality, 
which is in this way illustrated.
%
\subsection{Information System and Database: a formal approach}  
\label{ss:SIBDApprocheFormelle} 
%
\noindent
First let us point out that~:
{\sl databases ({\em \thinspace DB\thinspace}) 
are to information systems
what are computer programs to the intuitive notion of algorithm:
a formal, mathematical presentation
and the ability of an also formal processing}.
Indeed, algorithms and information systems are generally expressed
in a natural language  (in a more or less clear way)
and assume implicit content (which can be important)
and also unspoken comment (which may be quite a problem).
Recall that algorithms and information systems have existed since
Ancient Times\footnote{%
Some exhaustive descriptions of algorithms about trading and taxes
date from Babylonia (2000 BC to 200 AC).
Information systems really emerged with mecanography
(end of XIX~th century) and the development of computer science.
However, there are far earlier examples of what we could now call
information systems since they show a 
{\sl neat organization and presentation
of data on a particular subject\thinspace}:
for instance, the Roman census.
                  }.%
~In both cases, this formal expression is essentially done in the
framework of mathematical logic.
Observe that programming and algorithms are particularly related
to lambda calculus whereas databases and consequently 
information system are particularly related to set theory.
\medskip
\\
As concerns programs and algorithms, let us mention the remarkable
work of Yuri Gurevich
\cite{DershowitzGurevich--Article--Axiomatization-2008}.
He introduced a notion of Abstract State Machines ({\em ASM}),
which is based on model theory (in logic) and is a mathematical
foundation of the notion of algorithm 
which is as much as possible refined.
Not only does {\sl he captures the notion of algorithm,
but he also formalizes their operational mode}.
More precisely, Gurevich deals with {\em operational semantics},
i.e. {\sl the way algorithms and programs are executed},
(the outcome is the programming 
of an {\em interpreter\thinspace}
{\em and/or compiler\thinspace} and of an executor of programs).
This highly constructive operational point of view completes what is called
{\em denotational semantics} and which deals with what
{\sl algorithms and programs 
compute\thinspace}\footnote{%
Observe that these semantics correspond respectively to
Arend Heyting's semantics and
Alfred Tarski's semantics. 
                 }.%
\medskip
\\
This is, in fact, the way Gurevich states his thesis:
%
\begin{quote}
{\sl {\scriptsize$\ll$}~ASMs capture the
{\em \thinspace step by step\thinspace} 
of the execution of sequential algorithms}.~{\scriptsize$\gg$}
\end{quote}
%
For Gurevich, any given algorithm (in particular, any computer program)
``is" a particular ASM which is going to mimic his functioning.
This allows to consider an algorithm as a formal object
(namely, an ASM). Gurevich's thesis extends Church-Turing's 
thesis\footnote{%
\label{TheseChurchTuring}
Church-Turing thesis states that
"{\sl Every process or computation which can be done with a machine
in a purely mechanical way, i.e. all what is computable with a machine,
can be done with a Turing machine\thinspace}" (1936).
Thus, this thesis asserts that the intuitive notion of 
{\em effective computability\thinspace}
coincides with a formal mathematical notion: {\em computability
with Turing machines}.
This thesis was first stated by Alonzo Church (1932) with the model of 
$\lambda$-calculus (Church thesis)
which appeared at that time far more ``theoretical" than
Alan Turing machines, cf. the note~\ref{JLK}.
We shall look at the Kleene computability model
of recursive function (1936)  
in section~\ref{ss:BottomUpvsTopDown}.
A first  (complete) formal definition of recursive function was found
by Jacques Herbrand and formalized by Kurt G\"odel (1932).
                }%
~(at least for sequential algorithms):
indeed, Gurevich thesis proves it.
More precisely, Church-Turing thesis is about denotational semantics
(the diverse computation models which have been imagined are pairwise
equivalent: we say they are Turing-complete).
Gurevich extends this thesis to operational semantics:
ASM are a computation model which is {\em algorithmically complet}
(cf.~also section~~\ref{s:Conclusion} and 
\cite{FerbusZandaGrigorieff--Article--ASMLambdaCalcul-2009}).
What is really remarkable with ASMs is how 
their formalization is simple and natural,
which, in general, is not the case with the other approaches to
operational semantics of computer programs. 
We come back to ASMs (and their relation with Kolmogorov
complexity and classification) in the conclusion.
\medskip
\\
As concerns, information system (which is an intuitive notion) and
their modeling via database (which is a formal approach),
we shall see that, historically and conceptually, things were not as
simple as they were with programming and the formulation of
theoretical models for computability
--~which, indeed, occurred at a time when there was no computers.
In the case of information systems, it was all the opposite.
\medskip
\\
Recall that the first formalization 
of the representation and treatment of data,
(that is what is now called an information system)
is Codd's relational model for 
databases (1970) \cite{Codd1970}.
What was quite original with Codd's approach 
is the idea that there were
mathematics which should ``manage" information  in computers.
Though this may seem quite obvious now,
up to the time Codd created his theoretical model
(a time where programs were written on punched cards),
that was not the case:
computer files were stored in a great 
mess\footnote{%
{\em Multics\thinspace} was the first important operating system
to store files as nodes in a tree (in fact a graph).
Created in 1965, it has been progressively replaced since 1980
by {\em Unix\thinspace}. Derived from Multics, it includes a new
feature: multiple users management.
Now, all operating systems are based on Unix.
Multics was a turning point in the problem of data storage:
until now, one speaks of {\em hierarchical model\thinspace}
and {\em net model}.
But, in fact, these ``models" have been recognized as models
only after Codd introduced the relational model!
Finally, observe that the graph structure of the Web also comes
from the organization of files with Multics.
                  }.%
\medskip
\\
One of the most fundamental and unprecedented feature of
Codd's relational model is the formalization of the notion of
{\em query\thinspace}. He founded this notion on a new calculus:
{\em relational algebra\thinspace} 
which is a kind of {\sl combinatory
logic with operators acting on 
tables\footnote{%
This combinatory logic has much to do with the programming
language {\em Cobol} created in 1959.
               }}%
~joined together: classical set theoretic 
operations ({\em union}, {\em intersection},
{\em cartesian product} and {\em projections})
and also new operations: {\em selection\thinspace} and {\em join}.
It turns out that the join operator is
really a fundamental one in logic.
Codd also develops a {\em normalization theory\thinspace} to handle
the very difficult problem of removing {\em redundancies\thinspace}
in information systems.
\medskip
\\
Surprising as it is, though Codd worked in an IBM research center,
he had to fight very
 hard\footnote{%
The dedication in his last book (\cite{Codd1990}, 1990) is as follows:
%
{\scriptsize$\ll$}~To fellow pilots and aircrew in the Royal Air Force  
duriring War~II and the dons at Oxford.   
These people were the source of my determination to fight for what I 
believe was right during the ten or more years in which government,   
industry, and commerce were strongly opposed to the relational        
approach to database management.~{\scriptsize$\gg$}~. 
%
}%
~to impose his views.
The first implementation of his model was not done by IBM
but by {\em Oracle\thinspace},
at that time a very small
company\footnote{%
Oracle is now a company worthing billions dollars.
                 },%
~which saw its exceptional interest and implemented it in 1980.
It is only a few years later that IBM also implemented Codd's model.
Now, all existing DBMS (database management systems) are based
on Codd's relational model.
Let us mention that databases are still largely underrated
though it could be so profitable in many disciplines.
But this is clearly not to last very long due to the dissemination
of digital information
(with an incredible speed, no one could have expected a few years ago).
\medskip
\\
There is another theoretical model for databases:
the {\em Entity/Relationship model\thinspace} due to
Peter Pin-Shan S. Chen \cite{Chen76}.
This is a formal approach to databases which essentially relies
on Codd's relational model but is more abstract.
In this model, a database is represented as a graphic
which looks like {\em flow charts\thinspace} used in the 60-70s
to modelize computer programs.
It is the source of the language 
{\em UML\thinspace} \footnote{%
UML ({\em Unified Modelling Language\thinspace}) is a formal language,
which is used as a method for modeling in many topics,
in particular, in computer science with databases
and {\em Object-Oriented Conception} ({\em OOC\thinspace})
--~in fact, this is the source of UML.
                              },%
~which is a system of graphic notations for modeling.
In our opinion, Chen's theoretical model is very deep
and it should still be the source of many important works.
Databases rested on the Entity/Relationship 
model deserve to be called
{\em conceptual databases}.
They constitute an abstract logical extension
of relational databases which should 
have a fundamental role in the future
as concerns information processing, classification and any
algorithmic information theory.
\medskip
\\
{\em  Object-Oriented Programming Concepts\thinspace}
are also inescapable in information processing and in the approaches
to classification.
Let us mention the {\em inheritance\thinspace} concept in the difficult
problem of concurrent access to data, i.e. when the same data is used by
several actors: attributes, processes, systems, users.
Another important concept from Object-Oriented Programming
is that of {\em event-driven programming}:
a particular value in the execution of a program or particular data
in a database trigger the execution of some (other) program.
\medskip
\\
Lastly, let us mention another theoretical model for databases:
the {\em deductive model\thinspace}
(also called  {\em deductive databases\thinspace}).
This is also a fundamental model. It mixes Codd's relational
model and the predicate calculus, bringing intensionality
(i.e. abstraction) to Codd's model through the {\em in extenso}
adjunction of first-order variables.
The query language for deductive database is {\em Datalog}.
It is a pity that the existing implementations of Datalog,
which work quite well, are only used in some research labs.
Currently, there is no ``real" deductive DBMS
(Database Management System) with the same facilities offered by
relational DBMSs.
This is quite surprising as information system, with the Web,
have taken such a huge impact.
\medskip
\\
One can also question why diverse theoretical models,
as fundamental as they are, can coexist with no serious attempt
to mix them.
Maybe, this is because database is a very recent discipline,
quite probably, this will happen in the near future.
We are working towards this goal with the notion of
conceptual databases\footnote{%
\label{cite:LogSI-BDR-BDC}
Ferbus-Zanda~M.
{\em Logic and Information System: Relational and Conceptual 
Databases}.  
In preparation. 
                  },%
~using logic as a foundational theoretical basis.
Consider the general problem of classification of information.
Database, with the diverse theoretical models described above,
constitute a formal approach to that question.
Especially with the notion of query which becomes a mathematical
notion (which, moreover, is implemented)
far more sophisticated than keywords.
In fact, queries generalize
keywords\footnote{%
One should rather say that keywords --~used with web browsers~--
constitute very elementary database queries
(of course, database queries are much older than the Web
which emerged only in the 90's).
                   }.%
\medskip
\\
Whichever theoretical model of database is used,
a fundamental primitive notion is that 
of {\em attribute\thinspace}
(which can be seen as formal keywords)
and different kinds of set groupings of attributes 
so as to make up the {\em relational schema} of a database.
This constitutes the wanted 
{\em formal classification} of the initially
unorganized data.
The relational schema of a database is the structural part of a database: 
its {\em morphology}.
So, in the relational model, a database is structured in
{\em tables\thinspace}. The names of the columns of a table are some
attributes for the database.
A line in a table describes an entity (from real world):
this entity is reduced to
the values (for that line) of each of the attributes of the table
(i.e. the names of columns).
There are relations between the tables of a database,
kind of {\em pointers\thinspace},
which follow some ``diagram" relying on the chosen relational schema
of the database.
\medskip
\\
The content of the tables constitutes the {\em semantics\thinspace}
(otherwise said, the current {\em content\thinspace}) of the database
at some particular time.
Each table is structured in columns and can also be seen
as a set of lines (the so-called ``tuples").
The number of columns is fixed
but the set of lines  varies along time.
Each line is a set of values: one value per attribute
(recall columns and attributes are the same 
thing)\footnote{%
Lines are usually presented as tuples but, conceptually,
this is not correct: in Codd's relational model there is no order
between the lines nor between the columns.
Codd insisted on that point.
In fact, conceptually and in practice, this is quite important:
queries should be expressed as conditions (i.e. formulas)
in the relational algebra, using names of attributes and of tables.
For example, it means that queries cannot ask for the first 
or twentieth line (or column).
                }.%
~This notion of line corresponds exactly to that of card in physical files,
(for instance, those used to manage libraries in pre-computer days)
or to the content of a punched card (mechanography).
\medskip
\\
For instance, suppose we have a table about authors of books in a
library which has the following attributes:
{\tt AuthorSurname}, 
{\tt AuthorName},
{\tt AuthorCountry}, 
{\tt AuthorTimes}.
The {\tt AuthorSurname} column will contain names
(such as {\tt Duras}, {\tt Sarraute}, {\tt Yourcenar},
{\tt Nothomb}, {\tt Japp}, etc.).
A typical line could be   
\linebreak 
$\bf{ \{}$\thinspace 
{\tt AuthorSurname}\thinspace{\bf .}\thinspace{\tt Duras}~,~ 
{\tt AuthorName}\thinspace{\bf .}\thinspace{\tt Marguerite}~,~ 
{\tt AuthorCountry}\thinspace{\bf .}\thinspace{\tt France}~,~
\linebreak 
{\tt AuthorTimes}\thinspace{\bf .}\thinspace{\tt XX~th century}\thinspace 
$\bf{ \}}$
or also the $4$-tuple
{\bf (}{\tt Duras}~,~
{\tt Marguerite}~,~
\linebreak 
{\tt France}~,~
{\tt XX~th century}{\bf )}
since the ordering of values in this tuple 
makes it possible not to ``explicit"
the associated attributes.
\medskip
\\
Queries allow to access these contents.
Note that the content of the tables evolves through time due to
{\em updates\thinspace} of information:
{\em adjunction, removal, modification}.
A database looks like the set of sheets of a spreadsheet
({\em Excel\thinspace})  augmented with links
between them that are managed through queries
(which spreadsheets cannot do, or in a very rudimentary and complex
way).
\medskip
\\
Thesauruses (cf.~note~\ref{note:Thesaurus}) are, in fact, databases.
The relational schema of such a database is the structure of the
considered thesaurus, otherwise said, the layout, the architecture
of the thesaurus.
The diagram of this database (which is a graphic representation of 
its relational schema) formally expresses this architecture.
It is clear that there can be several tables in this database.
For instance, in a thesaurus dedicated to the epistemology of mathematics,
there could be specific tables for mathematical logic, probabilities,
algebra, topology, geometry, functional analysis, differential calculus,
integration, etc. and other tables dedicated to mathematicians
(mentioning the concepts they introduced),
to philosophers, to historians of mathematics, etc.
\medskip
\\
Of course, the choice of such tables is completely subjective.
One could structure the database very differently, considering 
{\em synonymy, quasi-synonymy, connectivity,  
analogy, comparison, duality, contrast, etc.}
among the diverse words of the thesaurus.
The internal organization of a given table (the choice of the columns
i.e. of attributes)
depends on what one intends to do with the thesaurus and on the choices
already made for the diverse tables.
The contents of the tables are then constituted by all the words
put in the thesaurus.
\medskip
\\
Without definitions, the thesaurus is a kind of hierarchical
dictionary of synonyms, associations, etc.,
i.e., a structure on keywords.
To augment it with definitions, we insert them as contents
of the tables in specific columns.
In any case, let us stress that the relational schema of the
associated databases essentially relies on the ``association"
part of the thesaurus (indeed, its graph) and not on its
``definition" part.
Also, observe that it is the power of computers and databases
which makes it possible to build and use such complete
thesauruses. It would unrealistic to try a readable paper version
of a dictionary which would be at the same time
a usual dictionary and a synonym dictionary and would also give 
definitions\footnote{%
\label{DicoCirculaires}
In fact, any such ``complete" dictionary is necessarily circular:
a word $a$ is defined using the word $b$ which is itself defined
with other words themselves defined in the dictionary.
It requires some knowledge {\em external\thinspace} to the dictionary to
really grasp the ``meaning" of words.
Note that this incompleteness is more or less ``hidden''.
On the other hand, in a synonym dictionary, the structure essentially
relies on circular definitions.
This is less apparent with paper dictionaries: for a given word,
there will be only references to its synonyms.
However, with digital dictionaries, this {\em circularity} is really striking:
links ``carry" the reader to the diverse synonyms
and can be implemented with pointers.
                    },%
~but any good computer graphical user interface makes it possible.
\medskip
\\
Note that what is not explicitly represented as a table can be recovered
via some query.
For instance, if we decided a structure by discipline, one can obtain
all synonyms of a given word, whatever be the table of the thesaurus
database in which they have been inserted
(according to their associated discipline).
This shows that any particular choice of a structure for the database
leads to no disadvantage as concerns the usage of the database:
whatever grouping of information is wanted, it can be obtained via
some appropriate query.
This is ``hidden" to most users which have no idea of the internal
organization of the database. 
In general, one chooses a structure which makes easier the
elaboration of the schema of the database,
or an {\em optimized\thinspace} structure to get efficient executions
of queries
(recall there are database tables containing millions of lines).
Of course, the synonymy in question is relative to the closed
world of the database formalizing the thesaurus.
\medskip
\\
The result of a query in relational 
databases is a {\em view\thinspace}
which is structured as a table.
The only difference between a table 
and a view is that views are stored
in the RAM (random access memory) of the computer
(which is a volatile memory: it disappears
when the computer is turned off)
whereas ``real" tables of the database
represent {\em persistent\thinspace}
data which are stored on non-volatile 
memory: hard disks, magnetic tapes, etc.
Of course, one can nevertheless save a view.
\medskip
\\
Observe a very interesting phenomenon with this example:
the emergence of the notion of database.
Indeed, following the same approach, one can build a database
dedicated to epistemology of physics, of chemistry, of biology, etc.
and group these databases in 
a unique database in order to get a thesaurus
dedicated to epistemology.
One can also group epistemology with other disciplines. 
Clearly, one has to fix the {\sl wanted 
level of abstraction/refinement\thinspace} to
build the thesaurus (or, more generally,
a database) and what is the limit
to the considered subject.
This is one of the most difficult problems in modeling.
Any scientific activity goes along a particular answer to that problem.
\medskip
\\
This example leads to the following observation: 
in this paper, the notion of ``object" has not been much considered.
It is clear that the {\em hierarchical \thinspace}
character on a thesaurus relies on {\em inheritance\thinspace}
(a concept from OOC, cf. above).
It seems therefore necessary to add to Codd's relational model
some concepts of the object oriented  
approach\footnote{%
Codd was strongly opposed to any addition from the object approach to
the relational model. Indeed, the so-called ``First Normal Form"
(due to Codd) formally forbids the possibility of an attribute
structured as a list, a tree or a graph (which is exactly what OOC would do).
When he elaborated his model, this was a reasonable choice:
the object approach is quite destructuring while Codd's approach was
a structuring one.
Let us mention that Codd also opposed Chen's
Entity/Relationship model (nobody's perfect)!           
                },%
~which is what we try to do with conceptual 
databases\footnote{%
{\em Ibid}.~Note~\ref{cite:LogSI-BDR-BDC}.
                       }.%
\medskip
\\
If we consider the general case,  we observe that
{\sl the notion of query in databases is essentially dependent
on the structure of the database associated to the relational schema}.
Database queries are similar to Google queries with one big difference:
queries in relational database are written in a programming language
which is far more sophisticated than conjunctions of keywords
allowed in Google queries.
In all 
implementations\footnote{%
An implementation of Codd's relational model for databases
is a DBMS ({\em DataBase Management System\thinspace}).
Any DBMS includes an interpreter of the language SQL
(such an interpreter is, in fact, an implementation of Codd's
{\em relational algebra\thinspace}, the fundamental calculus in this
theoretical model).
                                }%
~of Codd's relational model for databases,
queries  are written in the programming language {\em SQL\thinspace}
({\em Structured Query Language}).
\medskip
\\
As with keywords, the choice of attributes and that of groups of attributes
in a database is completely subjective: this is {\em semantics} and this
semantics is formalized by the relational schema.
Once such choices are done and the relational schema is fixed,
the form of possible queries is somewhat constrained but, nevertheless,
it is possible to ask whatever is wanted.
This was argued above with the example of the thesaurus.
As for the Web, such a relational schema is 
absolutely impossible because
the Web is so fundamentally dynamic.
\medskip
\\
{\sl Observe that, at any step, we have with databases
a precise idea of the structure
we are working on (it is a mathematical object)
and extracting information out of such a structure is done in a
rigorous way, using the formal notion of query.}
\medskip
\\
Let us then notice that the result of a
query is {\em exhaustive\thinspace} relative to the
database we consider: we get exactly all objects in the base
that satisfy the query, {\em no more no less}.
Also notice that the information content of a (correctly formalized)
database is precisely known at any time
and the modifications brought to the base
(adding, removing or changing data)
is precisely controlled.
Of course, this is not the case when extracting information
from the Web with a search engine
and this is not the case either for large data banks
(in biology, medicine, cartography, etc.)
which have no solid mathematical foundation
as have relational databases neither in the structuralization
of data nor for the queries.
Databanks are indeed databases which are 
somewhat not well formalized (or somewhat ill).
In other words databanks can be really databases whereas
this is intrinsically impossible for the Web.
%
\subsection{Database and bottom-up versus top-down duality}  
\label{ss:BDetModesBUvsTD} 
%
%
\noindent
Let us now look at the elaboration and use of databases
in the perspective of bottom-up et top-down approaches.
It turns out that this is much the same as with keywords and
Google queries.
\medskip
\\
) 
The choice of the relational schema is done using
a bottom-up or top-down operational mode. In general,
both modes are used jointly (in fact, alternatively).
In the bottom-up mode, one uses the expected future content
of the database to build its relational schema (which will structure
this content). In the top-down mode one builds the
relational schema on considerations which 
are external to the future content. 
At first glance, using the bottom-up operational mode may seem 
paradoxical:
to use the content in order to structure it.
But this is not the case.
\medskip
\\
In practice, to build a relational schema for a given database,
one starts from some sketchy idea of the schema, represents it
as some graphic (top-down approach), then implements it
(this is programming work).
A kind of {\em prototype\thinspace} is thus obtained.
This being done, one fills the tables of the database with a few lines
(a ``set of data") to test the pertinence of the relational schema,
which may lead to adjust it (bottom-up approach).
And this may be repeated\ldots~
Recall that the content of a database is precisely what gives the
semantics of the database whereas the construction of the
relational schema is morphology (syntax).
With such a mix approach, one can build the {\sl morphological
(syntactic) part of the database via some access to a part of
the semantics of the database. And vice-versa}.
\medskip
\\
Thus, this approach, so seemingly paradoxical, is not so.
In fact, there are two true difficulties.
First, to delimit the scope of the information system which is
to be modeled, and this is done using the given
{\em specifications\thinspace}.
Second, to choose the right level of abstraction of each component
(attributes, tables,etc.).
\medskip
\\
2)
The choice and programming of queries comes next.
And the approach is bottom-up, top-down and mix:
this is similar to what we said about the elaboration of
the relational schema.
However, for quite complex databases, one may have to build
the schema and the queries more or less simultaneously:
we saw this with the thesaurus.
\medskip
\\
3)
Once the relational schema of a database seems adequate
and the main queries have been written down and programmed
(some of them testing the coherence of the base),
one can really fill the database and complete its content.
Queries can be added as wanted. But any modification to the
relational schema, even a seemingly minor one, can cause a great
damage when the size of the database is somewhat huge.
For instance, breaking an attribute  {\tt Artiste\thinspace} into
two attributes  {\tt Composer\thinspace} and {\tt Interpreter\thinspace}
in a music database.
\medskip
\\
4)
{\sl The content of the database can then be grasped through a
completely top-down mode using queries}.
This is why relational databases are such a breakthrough.
Huge quantities of data can be accessed from the outside 
in a completely rigorous mathematical way.
Thus queries can be viewed as questions to the DBMS in which the
query processor really behaves as an {\em oracle\thinspace}
(since its works is invisible to the user).
Of course, one can also follow a bottom-up approach:
browse the content of the database to find some wanted information.
Before Codd's relational model, this was, indeed, the sole possible
approach (excepted mechanography) with the old physical ``files"
such as index cards in large libraries: alphabetical (syntactic) sorts
caused no problem but sorting such files according to themes (semantic)
was a real headache!
%
\subsection{Classification and bottom-up versus top-down duality}  
 
%
%
\noindent
Let us summarize. Approaches to classification via
keywords or via Google queries (such as 
Google classification), databases
(whatever theoretical model is used)
have the same intrinsic nature. 
{\sl In the diverse phases of the elaboration,
especially with keywords and queries, one can follow a
bottom-up operational mode or a top-down one}
(and generally, both modes are used alternatively
in a {\em mix\thinspace} mode).
{\sl Queries obtained in that way then allow to grasp sets of texts in a
top-down mode (that is with no understanding of the meaning of the texts)
and classify them}.
\medskip
\\
{\sl The approach to classification using compression is entirely relevant to
the top-down mode}.
Observe that, for the classification using compression,
the framework is then {\em purely syntactical},
there is no use of any keyword or query which would convey some semantics
(for instance, that given by the chosen identifiers).
{\sl Thus, one gets information relative to texts without turning to their
semantics: simply compress and compute}. 
\medskip
\\
At first glance, this approach may seem somewhat ``miraculous":
one is able to classify information contained in texts
without getting into their contents and with no need to understand them.
On the contrary, in the previous approaches, one is lead to use
a bottom-up mode (though this is not absolutely needed) to build
interesting queries (and the relational schema in a database). 
Let us recall what we evoked supra: text compression is a highly
theoretical science and a simple, current-use algorithm such
as ``gzip" is the result of years of research.
Of course, in classification by compression, texts are not
chosen randomly!
However, for the next future, one sees no limit to the usage of
the above method to all information which is on the Web.
\medskip
\\
Considering the general problem of classifying information,
observe that {\em statistics\thinspace} constitute a particular case.
Usually, the statistical approach is top-down, computing
{\em correlation factors\thinspace}
to group objects and/or informations
and get a structure on them.
Indeed, Google and compression algorithm heavily use
statistics.
Nevertheless, one can also follow a bottom-up mode
with statistics or even mix these two approaches.
This will be seen bellow where we propose a
probabilistic version of the comprehension schema
(cf.~section~\ref{ss:schemaCompProba}).
%
\section {Set theory interpretation of 
Bottom-Up versus Top-Down duality}
\label{s:InterpEnsDualiteBUvsTD}
%
\noindent
Let us now look the different approaches to classification in the
perspective of the comprehension schema in Zermelo-Fraenkel
set theory {\bf ZF}.
A theory which can be viewed as one of the first formal mathematical
attempts to approach the notion of classification, sets being the most
rudimentary way to group elements.
As a matter of fact, Codd's relational model 
for databases relies on (naive) set theory,
which is not so surprising in the search of 
a formal structuralization mode.
\medskip
\\
Thus, the {\em bottom-up versus top-down duality}\enskip
that we point in classification 
(cf.~section~\ref{s:ClassifInfoAppBUTD}),
can be illustrated by the way the set theoretical comprehension
schema ``works".
We also discuss a probabilistic version of the comprehension
schema which among others illustrates
the {\sl exact versus approximate\thinspace} duality.
%
\subsection{The set theoretical comprehension schema}  
\label{ss:schemaCompEns} 
\noindent
This is an approach from ``pure'' mathematics.
\medskip
\\It is a global approach, intrinsically deterministic,
going along a fundamental dichotomy:
%
\begin{center}
True/False,
\\
Provable/Inconsistent.
\end{center}
%
A quest for absoluteness based on {\em certainty}.
This is reflected in the classical comprehension 
schema
$$
\forall x\ \exists y~~~ y=\{z\in x 
\mbox{~~;~~} \mathcal{P}(z)\}\footnote{%
More formally: $\forall x~ \exists y~~\forall z~~  
(z\in y ~~ \longleftrightarrow~~(z\in x ~\wedge~ 
\mathcal{P}(z)\thinspace))$.
                                      }%
$$
where $\mathcal{P}$ is a {\sl known property fixed in advance}.
Thus, the set clustering is done from a well known property
which is defined within this dichotomy.
{\sl To do such a grouping and build such a set, we again find ourselves
in top-down operational mode: this set is being constructed from
the property $\+P$}.
\\
More precisely, with a constructivist approach:
%
\begin{itemize}
\item [\textbullet] 
We start with a set $x$. 
\item [\textbullet] 
We choose a property $\+P$ relative to elements of the set $x$.
This can be done in both bottom-up and top-down modes exactly as
in the choice of keywords for a query or as in the elaboration of a
query in a relational database (cf.~section~\ref{ss:BDetModesBUvsTD}).
Note that the idea of the grouping,
i.e. the {\em choice\thinspace of this grouping
(formalized by the property $\+P$}) is completely subjective: 
this is {\em semantics}.
Nevertheless, we can also get such a property $\+P$
in a syntactic way: through a computation
(cf.~section~\ref{ss:KBDIntentAbstSemant}).
\item [\textbullet]
Having this property $\+P$, we then pick the elements of the set $x$
which satisfy $\+P$. 
\end{itemize}
%
The comprehension
schema\footnote{%
One can also constraint in different ways this property $\+P$.
In particular, to avoid circularities such as the one met when
$\+P$ contains some universal quantification on sets, hence
quantifies on the set it is supposed to define
(this was called {\em impredicativity\thinspace} by Henri Poincar\'e).
                                  }%
~allows us to consider such a set construction 
(in the ZF axiomatic set theory).
\mbox{}\medskip
\\
If we do not relativize this construction to some fixed set $x$
(or, equivalently, if we consider a set containing all sets) then
we face Russel's paradox\footnote{%
Russel's paradox insures that the following extension of the
comprehension schema is contradictory:
$\exists y~~~ y=\{~z \mbox{~~;~~} \+P(z)~\}$, i.e.
$\exists y~~\forall z~~  
(z\in y ~~ \longleftrightarrow~~\+P(z))$.
Indeed, consider the property $\+P$ such that
$\+P(u)$ if and only if $u \notin u$, then we get
$y \in y$ if and only if $y \notin y$.
                               }.%
~Observe that the solution to this paradox really makes
sense: in this approach, one should start
from something, and it will be
from an existing set of objects to work with such a property!
Indeed, the elaboration of the property $\+P$
is made in a mix mode (as with queries in a relational
database) then we can start with a certain idea for the
property $\+P$ (related to what is the set $x$)
then ``pick" some elements in the set $x$ to get a better idea of $\+P$,
and then pick again some elements in $x$ and adjust $\+P$,
and so on.
\medskip
\\
Once this property has been ``set up"
(maybe getting it {\em in extenso}), one is now able to group
all elements of $x$ which satisfy $\+P$.
Of course, in the mathematical literature, no one present such
successive approximations to get a property:
the obtained property is given directly!
Nevertheless, this is how things are being done in general.
Computer scientists are used to such practice:
a modular approach is used to perfect a database or a program.
Of course, so do the mathematicians quite often.
\medskip
\\
It is important to note that the grouping, that is, the definition
of the set $y$ or its constitution
(though some would rather consider an explicit construction)
can be done in a {\sl top-down operational mode which is an
intensional mode.
Intensionality, (one can also say abstraction)
is expressed by that property $\+P$}.
This property plays the role of a question which leads to an 
instantiation of comprehension schema which really behaves as an
{\em oracle}. The answer of the oracle is exactly the set of 
elements of $x$ satisfying property $\+P$.
This is the the opposite (the dual in fact) 
of an {\em extensional description\thinspace}
(which gives the element, one by one) which is necessarily
done in a bottom-up mode.
\medskip
\\
Knowing in advance the property $\+P$ is a very particular case
which does not happen in most ``real" situations.
Below, we develop this aspect by proposing 
a ``probalistic'' comprehension schema.
Then we show in section~\ref{s:InformationIntentionnaliteK},
how this probabilistic schema can be generalized using 
Kolmogorov's complexity. 
This brings us to the relation between the algorithmic
information theory and classification which are the heart of this work. 
%
\subsection{The probabilistic comprehension schema}  
\label{ss:schemaCompProba} 
\noindent
In the probabilistic approach, much more pragmatic
than the logical one, 
{\sl uncertainty} is taken 
into consideration, it is bounded and treated 
mathematically\footnote{%
We refer the reader to William Feller \cite{feller}
and also Kolmogorov, \cite{kolmo33,kolmo83},
and Chaitin \cite{chaitin69}.
                      }.%
\medskip
\\ This can be related to a probabilistic version
of the comprehension schema where the truth of $\mathcal{P}(z)$ 
for instances of $z$
is replaced by {\em some limitation of the degree of uncertainty 
of the truth} of $\mathcal{P}(z)$.
Formally, together with $z$,
we have to consider a new parameter in $\mathcal{P}$,
namely the event $\omega$ of some probability space $\Omega$
and we have to fix some confidence interval $I$ of $[0,1]$
(representing some prediction interval).
Denoting by $\mu$ the probability law on $\Omega$,
the probabilistic comprehension axiom for property $\mathcal{P}$ now states
$$
\forall x\ \exists y\ ~~~ y=\{z\in x \mbox{~~;~~}
\mu(~\{\omega\in\Omega\mbox{~~;~~}\+P(z,\omega)\}~)\in I\}
$$
As was the case with the set theoretical comprehension schema,
{\sl one gets in a top-down operational mode to do such a grouping
and build such a set\thinspace} from property $\+P$ and interval $I$.
This is so even if we allow some degree of uncertainty for the truth or
provability of property $\+P(z)$
(which is then replaced by ~$\mu(\{\omega\in\
\Omega\mbox{~~;~~}\+P(z,\omega)\})\in I$\thinspace)
for particular instances of $z$.
\medskip
\\
Once again, this is a precise particular case: 
though its truth has some uncertainty, this property
{\sl is well defined and fixed in advance,
together with the confidence interval $I$}. 
However, such a schema is closer to many situations met
in the real world. As in the previous case, such a property $\+P$
(and the confidence interval $I$) allow to {\sl define the set $y$
in a top-down operational mode, that is to get an intensional, abstract
description of the set $y$}.
It is natural to consider as above, an underlying 
{\em oracle\thinspace} (the probabilistic comprehension schema),
which, given some property $\+P$ and interval $I$,
returns the set $y$
with non totally accurate answers 
(the interval $I$ limiting the inaccuracy).
Observe that, as above, the {\em choice\thinspace} 
of $\+P$ and $I$ is relevant to {\em semantics}.
Remark that there are other ways to formulate a probabilistic
comprehension schema.
\medskip
\\
As concerns groupings of information relevant to a
purely top-down mode (the grouping itself,
the elaboration of a property to do it,
the definition of sets of information),
we treat it in the next 
section~\ref{s:InformationIntentionnaliteK}
about intensionality and Kolmogorov complexity.
\medskip
\\
Let us simply recall
(cf.~section~\ref{ss:ClassificationetModesBUvsTD} )
that classification by compression and some methods based on
statistical inference allow to have 
such purely top-down approaches.
The particular of Google classification is exactly the same as that
of set theoretical and probabilistic comprehension schemas
(for Google, keywords play the role of a property $\+P$)
and that of classification via databases, up to one significant
exception: {\sl with Google, everything is moving:
answers as well as the keywords proposed in queries}.
\section{Information, intensionality , abstraction and Kolmogorov complexity}  
\label{s:InformationIntentionnaliteK} 
%
\subsection{Classification, database, intensionality, 
abstraction, semantics and algorithmic information theory}  
\label{ss:KBDIntentAbstSemant} 
\noindent
We stressed in section~\ref{s:ClassifInfoAppBUTD}
the importance of the Web expansion and the huge interest of
classification by compression and Google classification.
The Web can be seen as a gigantic {\em expert system}:
first, it is a huge information system
(this is the network aspect, software and hardware,
between machines and servers),
second, machines are used and programmed by human beings
(their brains) with far more intelligence than what is done
in the syntactic world of machines which can only compute.
\medskip
\\
Classification by compression (and Google classification)
will surely be more and more used with information on the Web.
The same is true with statistical inference methods.
In some sense, all these approaches are tightly correlated and,
as any approach to classification
(cf.~section~\ref{ss:ClassificationetModesBUvsTD}
and section~\ref{s:InterpEnsDualiteBUvsTD}),
they lead to {\em top-down approaches to information}.
In particular, {\sl they can be used
to grasp the information content
of a text (and more generally of a set of texts)
with no access to it ``from the inside", i.e. without
reading and understanding the text}.
These methods look for analogies with other texts,
the meaning of which is known,
or they compare their respective information content.
Somehow, they are ``profilers'' which will become incredibly efficient
in the near future when applied to information on the
Web\footnote{%
Recall that once an information has been put on the Web, it is almost
impossible to remove it\ldots
                  }.%
\medskip
\\
However we have also explained how these methods still lack some formal
development, in particular for the notion of query:
for any classification of information, the first question is to find back
information from this classification.
It is a fact that the notion of query to the Web 
(with Google or any browser)
is still not really formalized.
\medskip
\\
We have seen that Codd's relational database model led to a
completely mathematical structure and processing of the information
contained in computer files through the relational schema
and the possible queries to the database
(the scope of such queries being tightly dependent of the 
relational schema).
As said above, before Codd, there was no such 
information processing with machines.
Codd had to fight to impose his mathematical model
and, even today, operating systems do not really use databases. 
A reflexion about possible formalizations of
classification by compression, Google classification
and a notion of query to the Web,
is, in our opinion, quite fundamental.
Note that with Google (or any other browser)
we have no idea how to measure the degree of uncertainty of
Google's answers.
The percentage of pertinent answers may be anything between
0\% and 100\%.
Google answers are {\em unpredictable} and {\em constantly moving}.
Not an easy situation!
However, it seems reasonable to ignore 
at first the moving character of
Google (and also its not completely scientific features,
cf.~section~\ref{ss:DiscussionMethode}, point~4)
when looking for a mathematical modeling of these methods.
\medskip
\\
Indeed, one starts from a clustering or 
more generally from a classification,
obtained by way of conjunctions of keywords which are proposed into 
queries for Google or from a clustering 
or a classification obtained by 
compression or observed by way of the statistical methods.
\medskip
\\
{\sl In the simple case of a clustering,
we infer the existence of a property, of a ``law'', 
which is a form of regularity.
The emergence of such a law coincides with the 
existence of a certain {\em degree of 
intensionality} in the clustering we accomplish.
Otherwise said, we make obvious a grouping 
of objects, the description of which
can be compressed by using this property. This is an intensional 
description (when the compression have been performed).}
{\sl This can be seen as an (extended) top-down
version of the set theoretical
or probabilistic comprehension schema:
the property used in the set groupings 
is not known and fixed beforehand}. 
\medskip
\\
For more sophisticated classifications, one will have 
{\em higher order clusterings}\thinspace, 
i.e. clusterings of clusterings, etc.
Otherwise said, several properties will be involved
(in some cases, even infinitely many properties, 
in a theoretical point of view).
Observe that, with a subtle analysis of modelization
using relational databases one can see that, up to now, 
quite a few levels suffices to modelize
a lot of discrete information systems (for the ``real world)''.
One can expect a similar situation for classifications obtained via the
top-down approaches as evoked above, at least for those relative to
the present real world.
\medskip
\\
In case of some random grouping, no law gives any description:
no classification is possible.
{\sl The sole descriptions which can be given are the 
extensional ones (element by element): 
they are intrinsically non intensional.
Such random groupings can be called ``non intensionalizable", in other
words, there is no shorter description and no more abstract one,
hence no more intensional one, which is equivalent}. 
Otherwise said such a description is incompressible.
\medskip
\\
This points out the remarkable pertinence of Kolmogorov complexity
theory which is an avant-garde theory. Especially when being considered with 
several points of view, namely by studying the randomness
of a word or its information content or the possibility to
compress this word.
Somehow, randomness is the ``opposite'' of classification,
More precisely, there is a duality  
{\em randomness versus classification},
coming from the fact that Kolmogorov's 
theory of algorithmic information
allows to look at these two sides of information
(this is what Kolmogorov explicitly tells in \cite{kolmo65}).
\medskip
\\
This duality is a quasi-opposition
though randomness is not {\em chaos\thinspace} (cf.~Part~I).
This points out deep relations between Komogorov complexity and 
relational databases (which constitute, up to now, as we saw, the sole 
implemented -- and widely spread -- logical approach
to information systems).
This complexity also appears unavoidable as soon as one is interested in
classification problems.
This is not surprise since Kolmogorov complexity is primarily
a theory about information!
\medskip
\\
If we go back to Kolmogorov's approach, one can observe that it is
relevant to the {\em top-down mode}.
Indeed, look at the basic definition of Kolmogorov complexity:
\begin{quote}
{\sl The length of the shortest program which outputs a given data}
(the output being a binary word which represents a given 
object)\footnote{%
$K_\varphi(y)=\min\{|p| : \varphi(p)=y\}$
where $K_\varphi : \+O \to \N$
where $\varphi:\words\to\+O$ is a partial function
(intuitively $\varphi$ executes program $p$
as a LISP interpreter does)
and $\+O$ is a set endowed with a computability structure.
We take the convention that $\min\emptyset=+\infty$
(cf.~Part ~I). 
                }.
\end{quote}
%
{\sl Larger is the Kolmogorov complexity of an object,
larger are all programs to produce it,
more random it is,
larger is its information content,
Larger is the Kolm are all programs to produce it,
less intensional is any description of it,
less intensional is it itself,
less abstract is any property that allows us to describe the object
(when we consider the property in a syntactical perspective) .}
\medskip
\\In this definition one does not enter into the content of the output
or into the details of the object, which is therefore taken as a whole.
{\sl One solely handles the object from the outside} via 
{\em some program and/or some property which allows to describe it.}
This is indeed a top-down approach as are classification using 
compression, classification using Google and 
a part of statistical inference methods. 
And this suggests that these classifications methods are somewhere related
and that Kolmogorov complexity could give an unifying 
mathematical formal framework.
\medskip
\\
In other words, thanks to Kolmogorov theory,
we are able to measure the complexity
of an object (in the sense of Kolmogorov), i.e. to give
a numerical measure of the {\em degree of
intensionality\thinspace} or even of {\em degree of
abstraction} which is contained in
a computable description of that object.
It is remarkable that this can be done
with no prerequisite ``knowledge'' of the structure of the object
and that this is indeed what allows us to apprehend this structure.
%
\subsection{Kolmogorov complexity and 
information theories, semiotics}  
\label{ss:KTheorieInfoCommSemiotique}  
%
\noindent
Let us now compare the diverse ways 
to approach the notion of information 
followed by Shannon (cf.~Part~I),
Kolmogorov, Codd and other researchers.
\medskip
\begin{itemize}
%
\item[\textbullet]
For Shannon (1948)
\cite{shannon48},
an information is a {\em message\thinspace} 
which is transmitted through some physical device.
In particular, an information is a signal and there can be losses
during the transmission.
This design is that of a {\sl dynamic information approach~}
and the physical communication medium is of outmost importance.
\medskip
\\
So he looks at robustness of information and comes to
a quantitative notion of information content in transmitted messages.
To measure variation of this quantity, he borrows
to thermodynamics the concept of {\em entropy\thinspace}
and he bases his theory on it.
So he clarifies, on mathematical basis, how to deal
with noisy communication channels.
In Shannon's theory, words represent information (messages). 
It is based on coding letters or groups of letters in a word
(cf.~Partie~I), i.e. {\sl it is a purely syntactic analysis of words
(and messages they represent) which makes no use of any semantics}.
\\
Thus Shannon elaborates a mathematical theory of 
the information content of messages transmitted with 
some loss of signal.
Its main (and hugely important) applications 
are related to telecommunications
(no surprise: Shannon worked in Bell Laboratories).
%
\item[\textbullet]
The origin of Shannon's work is Wiener's cybernetics
(cf.~note~\ref{BoiteNoire}) in the late 40's.
This subject was much discussed in the Macy
conferences (New-York, 1942~--~1953),
to which Shannon attended.
Before Wiener and these conferences, there was nothing like an
information theory.
\\
Cybernetics is a theory which establishes, among other things,
the concept of {\em auto regulated system\thinspace},
in terms of~: {\em global behavior}, {\em exchanges}, 
{\em communication} and {\em interactions}.
Fundamentally, this is a top-down approach to information and 
systems.
Wiener talks about~ ~{\scriptsize$\ll$}~{\sl a science of 
relations and analogies between (living) organisms
and machines\footnote{%
Wiener's book {\em Cybernetics or Control and Communication 
in the Animal and the Machine\thinspace},
published in 1948, raised many controversies
(and Wiener exchanged a lot with von Neumann about it).
                     }~{\scriptsize$\gg$}}.%
~In particular, he studies {\em random processes\thinspace} 
and the {\em ``noise''\thinspace} occurring during the exchanges
in a system.          
A fundamental notion in his theory is that of
{\em feedback\thinspace}~: {\scriptsize$\ll$}~An object is controlled
by the instantaneous error margin between its 
assigned objective~{\scriptsize$\gg$}.
This is clearly a prefiguration of Shannon's information theory
(Shannon attended Wiener lectures as a student).
\\
Wiener has an avant-garde vision on {\em machines}\thinspace!
His works are the origin of many 
discoveries, in particular, in sociological,
psychological and biological aspects of {\em communication\thinspace} 
and {\em interaction\thinspace} 
and, more generally, in all information theories.
Besides several research themes generated by Wiener's theory,
let us also mention that Wiener's theory has a deep influence on
a large part of
modern {\em semiotics}\footnote{%
A subject going back to Charles Sanders Pierce (1839 - 1914).
                         }.%
%
\item[\textbullet]
In particular, Let us cite Umberto  
Eco\footnote{%
Eco is President of the ``Scuola Superiore di Studi Umanistici",
University of Bologna, where he holds the chair of Semiotics.
He published many novels, essays and academic texts
in which he puts into practice his theories on semiology and language.
             },%
~in {\em The Open Work\thinspace}\footnote{%
\label{cite:Eco--OeuvreOuverte}
Eco~U.
{\em The Open Work}.
Bompiani, 1962 \& Harvard University Press, 1989.
                }%
~(1962)
which analyses the question of {\em openness\thinspace}
of art pieces
(that we can see as some form of non-determinism or as a plurality
of interpretations).
Eco often refers to Wiener in chapter~3~:
{\em Openness, Information, Communication}.
He convincingly pinpoints the necessity to 
distinguish between\footnote{%
{\em Ibid}.~Note~\ref{cite:Eco--OeuvreOuverte}.
                   }~:%
%
%
\begin{quote}
{\scriptsize$\ll$}~[\ldots] the {\em 
signification of a message\thinspace} 
and {\em the information\thinspace} 
it brings.~{\scriptsize$\gg$}
\end{quote}
%
%
In other words, it is important to differentiate
the {\em semantics\thinspace} of a message 
and its information content.
Eco gives a simple and illuminating example 
(which we slightly modify)
to make clear this distinction: the message 
``tomorrow it will snow in Paris"
does not have the same meaning in December than in August!
He also adds:
%
\begin{quote}
{\scriptsize$\ll$}~Wiener said that signification 
and information are  synonyms,
both related to entropy and disorder.
[\ldots information also depends on the source which sends
he message.~{\scriptsize$\gg$}
\end{quote}
%
%
Otherwise said, contrary to Wiener (and Shannon),
Eco stresses how the information content of a message
(and somehow its {\em pertinence\thinspace} too)
depends on the {\em context\thinspace} in which the message
is considered. 
We shall see below how Kolmogorov answers this problem.
%
\item[\textbullet] 
For Kolmogorov (1965) (see also Chaitin (1966) 
and Solomonoff (1964)),
the fundamental aspect of information is the
{\em information content} of an object, 
independently of any consideration on how this information
is used (as a message for instance).
This is a {\sl static vision of information}.
\medskip
\\
What Kolmogorov is interesting in is to give mathematical
foundations for the notion of {\em randomness\thinspace}
and to explicit the notion of {\em information
content\thinspace} of a given object which is 
{\em intrinsic\thinspace} to that object.
Thus, what Kolmogorov looks for is a
mathematical theory of information
which would be far more abstract than Shannon's one
and would be  based on semantics 
not only on a ``physical" object like a word.
His solution is to consider computer 
programs (considered as computable descriptions)
--~considering things in fact in the 
context of the calculability theory~--
which output an object and look at the 
length of a smallest one.
Thus, considering both programs and what the program does,
the algorithmic information theory created by Kolmogorov has both
syntactic (length of a program) and semantic 
features (i.e. what the program does).
\medskip
\\
With Kolmogorov complexity, one can 
capture an ``objective" mathematical
measure of the information content of an object. 
Moreover, this measure is really inherent to the object
--~ in some way it is an {\em universal} specification of the 
information content of of the object~--
since it does not depend (up to a constant) on 
the considered programming language to
get programs: this is the content of 
{\sl Kolmogorov's Invariance Theorem}.
In order to aim an ``absolute''
mathematical notion of randomness, 
Kolmogorov makes a drastic abstraction from any physical device
to carry information.
In this way, he elaborates the algorithmic information
theory which allows to 
``compute''\footnote{%
Recall that the very original idea on which 
Vitanyi based the classification
using compression is to compute an 
{\em approximate value\thinspace}
of this complexity via usual compression algorithms.
                     }%
~Kolmogorov complexity of any object.
Introducing a conditional version of Kolmogorov complexity,
he refines this notion of intrinsic complexity of an object by
{\em relativizing it to a context\thinspace}
(which can be seen as an {\em input\thinspace} 
or an {\em oracle, etc.\thinspace} for the program)
carrying some extra information.
This exactly matches the problem pointed by Eco
about the necessity to distinguish signification and
information content.
\medskip
\\
This is how Kolmogorov founds
{\em algorithmic information theory},
which can be looked at as much as a 
{\sl mathematical foundation of 
the notion of randomness}
than as a {\sl mathematical foundation
of information classification and
structuralization}.
%
\item[\textbullet]
As seen above, for Codd (1970), 
the fundamental feature of information is its
{\em structuralization\thinspace} --~which is formally described~--
and the fact that one {\em can get back information
from this structuralization in an exhaustive way}.
Codds theory essentially relies on mathematical logic.
Thus, Codd bases his work on the static aspect of information
Observe that, as Kolmogorov does,
Codd also makes abstraction of the physical device
carrying the information.
This was quite a revolution in information treatment at IBM:
previously, any information treatment dealt with
the {\em files\thinspace} containing the data:
information and files were considered as a whole.
\\
Observe that the modeling of information systems
via relational databases
also takes into consideration the subtle distinction raised by
Eco between semantics and information content:
{\sl the pertinence of an information with respect to a given
information system} is seriously considered.
The same distinction is taken into account in the construction
of the relational schema of a database.
For instance, in a database to manage a university,
a choice is to be made:
is the information about the students hobbies to be considered
or to be ignored?
Of course, this choice is completely subjective, this is semantics.
If the attribute {\tt StudentHobby} is retained then it will appear
in the relational schema of the database, i.e., in the
{\sl syntactic counterpart of what is retained as the 
``constitutional'' semantics of the information system}.
\end{itemize}
%
%
\subsection{Algorithmic information theory, representation 
and abstraction}  
\label{ss:ThAlgoInfoRep} 
%
\noindent
A priori, Kolmogorov complexity does not apply directly to the objects
we consider, but only to binary words associated to a chosen representation
of objects.
However, for the usual different representations, this has quite a minor
incidence (this is the content of the 
{\em invariance theorem\thinspace}).
Thus, we (abusively) speak of the Kolmogorov complexity of objects
instead of {\em Kolmogorov complexity of representations of objects}.
\medskip
\\
Nevertheless, if higher order representations are considered, this is no
more true.
For instance, if we represent integers 
as cardinals of (finite) recursively
enumerable sets. 
Indeed, Kolmogorov complexity allows to compare higher
order representations of integers, leading to a proper 
{\em hierarchy} of natural semantics for integers 
({\em Church iterators, cardinals, ordinals, etc.})
as we proved in \cite{ferbusgrigoKandAbstraction1}.
This hierarchy can be put in parallel with a hierarchy of 
Kolmogorov complexities obtained by considering infinite 
computations and/or oracles.
\medskip
\\We show, among other things, that 
Kolmogorov complexity is also useful to get a
kind of classification of semantics for integers which is rather 
amazing. We can also see this classification of different 
representations of integers as a {\sl 
classification of the degree of intensionality of 
these representations, i.e. a sort of classification 
of the less or more abstract
nature of different definitions of integers, obtained from the
different semantics we consider.}
We develop this in 
\cite{ferbusgrigoKandAbstraction1}\footnote{%
As in the two forthcoming (technical) papers: 
Ferbus-Zanda~M. \& Grigorieff~S. 
{\em Kolmogorov complexity and higher order set theoretical 
representations of integers} and 
Ferbus-Zanda~M. \& Grigorieff~S.~{\em Infinite computations, 
Kolmogorov complexity and base
dependency}.
                                              }.%
%
\section{Conclusion}
\label{s:Conclusion}
%
\noindent
The previous considerations show, in particular,
that not only Kolmogorov complexity
allows a mathematical foundation of the notion of randomness,
but this theory is also intrinsically 
related to the fundamentals of information:
the notions of {\em  information content\thinspace} 
and {\em  compression},
that of  {\em  classification\thinspace} 
and {\em  structure\thinspace},
and more generally, {\em database\thinspace} 
and {\em  information system\thinspace}
(as they currently are).
This theory is also related to the notions of
{\em\thinspace intensionality\thinspace}
and {\em  abstraction\thinspace}, and also
to the notions de {\em  representation\thinspace},
{\em  syntax\thinspace}
and {\em  semantics\thinspace}. An enormous scope!
\medskip
\\
This double aspect (randomness and classification)
--~drawn by Kolmogorov since the origin of his theory \cite{kolmo65}~-- 
is partly stressed by the denomination
{\em \thinspace algorithmic information theory\thinspace}
commonly used to distinguish Kolmogov complexity theory and
Shannon's {\em \thinspace information theory\thinspace}.
Many applications can be expected in various unsuspected domains.
And this theory seems to us particularly suited to provide a
{\em \thinspace unifying theoretical framework\thinspace} for a lot
of approaches to information processing.
\medskip
\\
However, it seems to us to be interesting,
to look for an extension of Kolmogorov complexity.
As it is now, it is essentially based on the theory of 
{\em\thinspace computable functions\thinspace}
hence on {\em\thinspace algorithms}.
What we propose it to extend it by considering
to {\em \thinspace sets}, {\em \thinspace information systems} 
and {\em databases}.
This would put forwards a  {\em\thinspace relational}, 
{\em\thinspace non deterministic} point of view
which would be in contrast with  the
{\em\thinspace functional}, essentially {\em\thinspace deterministic}
current point of view, first considered by Kolmogorov himself
(this goes along with a new look to ASMs in the relational framework).
It would then be possible to revisit (and to increase)
Kolmogorov complexity and ASMs in terms of the duality
{\em functional versus relational}
(see section~\ref{ss:BottomUpvsTopDown} et
section~\ref{ss:SIBDApprocheFormelle})\footnote{%
\label{cite:KetASMRelationnelles}
We study the duality of functional and relational in 
\cite{FerbusZanda--Article--DualiteLogComputerBooleanAlg}.
The relation between ASMs and Kolmogorov complexity
and the reconsideration of these theories with a
relational point of view are developed in a forthcoming paper:
Ferbus-Zanda~M.
{\em Kolmogorov Complexity and ASM: the relational 
point of view}, in preparation.
%
                                                }.%
%
\medskip
\\
This means that we look at Kolmogorov complexity
with a more refined and {\em more structured\thinspace} point of view
--~in other words with a {\em qualitative\thinspace} point of view~--
than that of Kolmogorov.
For him a program and an output are binary words
(which can represent sets, graphs, information systems, etc.)
and his main purpose is to get a  {\sl quantitative definition of the  
complexity of an object}.
\medskip
\\
Such a qualitative approach was also followed by Codd  himself
while he elaborated the relational model for databases.
His theory is based on the formal notion of attribute which is to
represent qualitative characteristics of objects
(which are related via diverse links which are also of qualitative
nature) and Codd puts such attributes in a mathematical framework.
A database is a formal and mathematical specification as ``scientific"
as any algorithm which processes data and computes.
\medskip
\\
In particular, one can look at the smallest program which outputs some
given object rather than at its sole length
or also look at the set of all programs 
which give the wanted output.
Such an approach enlightens new links 
between algorithmic theory of
and Gurevich's ASMs\footnote{%
This is what we started in
{\em Ibid}.~Note~\ref{cite:KetASMRelationnelles}.
                }.%
\ It opens promising perspectives.
As Gurevich told us\footnote{%
Personal communication while he was visiting our university in Paris.
                 },%
~the ideas of Kolmogorov complexity theory are far from having exhausted
all possible applications: it is just the beginning...
Classification of information by compression and Google classification
witness such new possibilities.
It is also in such a structural perspective that Bennet developed the
logical depth complexity \cite{bennett1988}
which considers the running time of the program
which gives the output.
It is also called the {\em organized complexity}.
\medskip
\\
Keeping the same spirit (with such a level of refinement),
comes this question: 
%
\begin{quote}
{\sl Why consider the shortest program?
What is so particular with it?}
\end{quote}
%
The answer comes from the observation of ASMs and
the Curry-Howard correspondence:
%
\begin{quote}
{\sl The shortest program is the most possible abstract.}
\end{quote}
%
Indeed, Curry-Howard correspondence insures a deep relation
between logic and $\lambda$-calculus 
--~in that sense, this correspondence is an isomorphism~--
hence by extrapolation logic and computer programming.
Curry-Howard correspondence plays a
fundamental role in the articulation of
proof theory, typed lambda calculus,
theory of categories 
and also with models of computing
(either theoretical or implemented ones 
like programming languages).
It was known by Curry for  combinatory logic 
as early as 1934 and  for Hilbert proof systems in 1958.
It was extended by William Howard in 1969 who
published a corner-stone paper\footnote{%
Howard W.
{\em The formulas-as-types notion of construction},
in {\em Essays on Combinatory Logic, Lambda Calculus and 
Formalism}.
Seldin J.P., Hindley J.R. eds., 
Academic Press, pp.~479-490, 1980.
                }%
\ in 80
\footnote{%
Joachim Lambeck also published in the 
70's, about this correspondence concerning 
the combinatories of the cartesian closed
categories and the intuitionist propositional logic.
Note that Nicolaas Debruijn ({\em Authomath system\thinspace})
and Per Martin-L\"of had also a decisive influence upon
the original Curry-Howard isomorphism. 
Martin-L\"of saw the typed lambda calculus, which he
was developing, as a (real) programming language
(Cf.~ Martin-L\"of~P.
{\em Constructive Mathematics and Computer Programming}.
Paper read at the 6-th International Congress for Logic,
Methodology and Philosophy of Science, Hannover, 
22 -- 29 August 1979.)
Similarly, Thierry Coquand elaborated the 
{\em theory of Construction}, on which is based the {\em Coq 
system}, iniatially developed by G\'erard Huet at
the INRIA (France) in the 80's. (See~ also note~\ref{JLK}).
                 }.%
\medskip
\\
Let us say briefly that in the Curry-Howard correspondence,
one consider that:
%
\begin{itemize}
\item [\textbullet]
Logical formulas correspond to types in typed $\lambda$-calculus
and to abstract types in computer science.
\item [\textbullet] 
Logical proofs correspond to $\lambda$-terms and computer programs.
\item [\textbullet] 
Cut elimination in a 
proof\footnote{%
\label{Coupure}
The notion of {\em cut in the Sequent Calculus and the Natural 
Deduction\thinspace} is a fundamental notion in proof theory. 
It was introduced by Gerhard Gentzen in the 30's -- and these two 
logical calculus too.
In some cases one can see a cut as a form of abstraction where
a multiplicity of particular cases are replaced by a general case.
In the sequent calculus, a cut is defined by means of the {\em cut 
rule}, which is a generalization of the {\em Modus Ponens}.
The fundamental result of Gentzen is the {\em Hauptsatz}, which
states that every proof in the sequent calculus can be transformed in 
a proof of the same conclusion without using this cut rule.
                }%
~corresponds to normalization by diverse
rules in $\lambda$-calculus, including 
$\beta$-reduction\footnote{%
\label{JLK}
In fact, Church's original $\lambda$-calculus 
can be extended with constants 
and new reduction rules in order to extend to classical logic
with the notion of {\em continuation\thinspace},
Thimothy Griffin, 1990.
--~and possibly classical logic plus axioms such as
{\em the axiom of dependent choice}~-- 
the original Curry-Howard correspondence between intuitionist logic
and usual typed $\lambda$-calculus.
This is the core of Jean-Louis Krivine's work who introduced some of those
{\em fundamental constants\thinspace} 
which have a deep computer science significance
(cf.~%
Krivine~J.L.
Dependent choice, `quote' and the clock.
{\em Theoretical Computer Science}.
308, p. 259-276, 2003.
see also: http://www.pps.jussieu.fr/$\sim$krivine/).
                             }%
~relating $\lambda$-terms and runs of computer programs. 
\end{itemize}
%
This enhances the abstract character of programs evoked above.
Indeed, {\sl the smallest logical proof (considered in a given context)
is in fact the one which contains the most numerous {\em cuts}}.
We saw (cf.~Note~\ref{Coupure}) that in some cases,
{a cut is a form of abstraction}.
Notice that a proof, of which we have eliminated cuts
(which therefore means in some situations replacing 
``a general case" by a lot of
``particular cases"), has its size bounded in the absolute by a ``tower of
exponentiations''\ldots
%
\begin{quote}
{\sl The more cuts a proof contains the more abstract it is.
Somehow, we can say that
the more abstract is a proof, the more compressed it is}.
\end{quote}
%
In the same way, 
%
\begin{quote}
{\sl The more redexes there is in a 
$\lambda$-term\thinspace\footnote{%
A redex in a  $\lambda$-term $t$ is a subterm of $t$ on which a one-step
reduction can be readily applied, for instance, with $\beta$-reduction,
this is a subterm of the form $((\lambda x. u) v)$
and it reduces to $u[v/x]$, which is the term $u$ in which every occurrence
of x is replaced by $v$ (some variable capture problems have to be
adequately avoided).
                                  },%
~The more abstract is a $\lambda$-term, the more compressed it is}.
\end{quote}
%
And for computer programs, the notion of cut can also be defined for
programming languages with their usual primitive instructions.
For instance, a program containing 
%
\begin{center}
{\tt for i = 1 to 1000000 do print(i)}
\end{center}
%
is more  abstract than the same program in which this loop
is replaced by the sequence of instructions
%
\begin{center}
{\tt do print(1) and do print(2) and \ldots~and print(1000000)}
\end{center}
%
Thus, the {\tt for} loop allows for cuts.
Hence a result similar to those precedents:
%
\begin{quote}
{\sl The more cuts a program contains, the more compressed it is}.
\end{quote}
%
\medskip
Observe that the more a program is compressed via cuts,
the more {\em declarative} is this program.
Which means that its text contains less {\em control} instructions,
i.e. less instructions about the technical way some parts of the
program are to be  executed.
A fully compressed program is totally declarative.
\medskip
\\
But what about ASMs in this context?
\medskip
\\
As we said, ASMs allow to represent- in a very simple way -
the step by step of the execution of any sequential algorithm
using models in first-order logic and some simple primitive instructions.
As can be expected, it is interesting to look for a notion of cut in the
ASM framework.
In the same vein, deep relations exist between ASMs, $\lambda$-calculus
and Curry-Howard correspondence.
Cf. our paper to appear in honor of Yuri Gurevich
\cite{FerbusZandaGrigorieff--Article--ASMLambdaCalcul-2009},
in which we represent ASMs in $\lambda-calculus$, showing that
$\lambda-calculus$ is {\em algorithmically complete}
as are ASMs. 
\medskip
\\
Going back to Kolmogorov complexity, 
we can can say that:
%
\begin{quote}
{\sl The shortest program producing a 
given output is the most abstract one,
hence (viewed in $\lambda$-calculus) it is the $\lambda$-term
containing most redexes, hence also (viewed in proof theory)
 the proof which contains the most
cuts}.
\end{quote}
%
In any case, this is a form of abstraction. Which is no surprise since
we already noticed that Kolmogorov complexity is fundamentally related
to the notion of abstraction.
\medskip
\\
Going back to the information context, 
we can say : 
%
\begin{quote}
{\sl Knowledge is abstract information: abstract, compressed,
with some intensionality content}.
\end{quote}
%
And such a knowledge will be, in its turn, compressed, etc.
This is exactly the mode the brain functions with language and mathematics.
Observe that some abstractions are somewhat ``accidental":
they occurred at some time and drastically modify the state of knowledge.
Such an abstraction was the invention of {\em phonetic transcription} of
Indo-European languages: with a handful of symbols
as (letters of the Roman or Greek alphabet and some extra signs),
one can write down all texts in these languages.
One can also enunciate them (which is not the same as
understanding them): a few rules suffice to
capture specific pronunciation features in any
such language.
Such an abstraction is lacking in Chinese writing\ldots
\medskip
\\
Note that it is really what Kolmogorov complexity shows.
Suppose an integer has a long and 
seemingly lawless binary (or decimal)
representation: it takes space to represent it in this way.
But if we get a (good) constructible property about this integer, then
we can obtain a short, abstract, compressed characterization of it.
And this increases our knowledge.
In the same way, the development of integral calculus,
some parts of geometry and fractal geometry,
allow for short (effectively computable) sequential descriptions
of shapes.
\medskip
\\
Especially, it appears that Kolmogorov's complexity
can be a very useful theory in order to address in a 
mathematical way the approaches of  classification,
which are now essentially, to the exception relational database,
heuristic methods (not yet fully formalized 
as can be expected from a classification method) such 
as classification using compression and Google classification. 
One can also hope for applications in other domains such as
semiology, cognitive science or biology with the genome,
as spectacularly shown by the French biologist Antoine Danchin in
his book, 
\cite{Danchin--Livre--BarqueDelphes-VAA-1998-2003}.
Indeed, classification by compression is already used by some 
biologists in such a perspective.
\medskip
\\
Let us conclude by stressing again
how much useful are such classification methods
using compression or using Google
along the top-down operational mode.
In many cases, we face huge families of objects 
(when one can define them) for which there
is no obvious structure.
{\sl So that we really are in a  syntactic
world and want to grasp this world with some semantic}.
This is, for example, the case for DNA sequences of living
organisms and for the multi billion many files on the Web\ldots
\medskip
\\
For that last example, though we are not so much pessimistic,
let us cite Edsger W.~Dijkstra's penetrating analysis in his famous 1972 
Turing award reception speech 
\cite{Dijkstra1972}\footnote{%
Let us mention the remarkable collection of Dijkstra's unpublished
papers and notes \cite{Dijkstra1982}.
                           }~:%
\medskip
\begin{quote}
{\scriptsize$\ll$}~As long as there were no machines, 
programming was no problem at all;
when we had a few weak computers, programming became a mild problem,
and now that we have gigantic computers, programming has become
an equally gigantic problem.
In this sense the electronic industry has not solved a single
problem, it has only created them -- it has created the problem
of using its products.~{\scriptsize$\gg$}
%
\end{quote}
%
%
\medskip
\mbox{}\\{\bf\large Remerciements.}                  \nopagebreak[4]
\medskip
\\
%
{\sl For Francine Ptakhine, who
gave me liberty of thinking and writing}.           \nopagebreak[4]                                           
\\
{\sl Thanks to Serge Grigorieff and Chlo\'e Ferbus  \nopagebreak[4]
for listening, fruitful communication
and for the careful proofreading                 \nopagebreak[4]
and thanks to Maurice Nivat
who welcomed me at the LITP in 1983.}
%
%

%
\end{document}